\documentclass[12pt]{article}
\usepackage{amssymb,amsmath}
\usepackage{graphicx}
\newcommand{\ZZ}{\mathcal{Z}}
\newcommand{\WW}{\mathcal{W}}

\newcommand{\VV}{\mathcal{V}}

\newcommand{\be}{\begin{equation}}
\newcommand{\ee}{\end{equation}}
\newcommand{\bea}{\begin{eqnarray}}
\newcommand{\eea}{\end{eqnarray}}
\newcommand{\beas}{\begin{eqnarray*}}
\newcommand{\eeas}{\end{eqnarray*}}
\newcommand{\ba}{\begin{array}}
\newcommand{\ea}{\end{array}}

\newcommand{\tr}{\mathrm{Tr}}

\newcommand{\is}{\! &\! = \! & \!}

\newpage

\renewcommand{\sp}{\hspace{1pt}}

\newcommand{\ccc}{\mbox{\large $ c$}}

\def\identity{{\rlap{1} \hskip 1.6pt \hbox{1}}}
\newcommand{\nbox}{{\,\lower0.9pt\vbox{\hrule \hbox{\vrule height 0.2 cm \hskip 0.19 cm \vrule height 0.2 cm}\hrule}\,}}

\def\href#1#2{#2}

\makeatletter
\@addtoreset{equation}{section}
\makeatother

\def\appendix#1{
  \addtocounter{section}{1}
  \setcounter{equation}{0}
  \renewcommand{\thesection}{\Alph{section}}
  \section*{Appendix \thesection\protect\indent \parbox[t]{11.15cm}
  {#1} }
  \addcontentsline{toc}{section}{Appendix \thesection\ \ \ #1}
  }

\newcommand{\deltaa}{\delta_{1}}

\begin{document}
\begin{titlepage}
\hfill
\vbox{
    \halign{#\hfil         \cr
           } % end of \halign
      }  % end of \vbox
\vspace*{15mm}
\begin{center}
{\Large \bf A Massive Study of M2-brane Proposals}

\vspace{1.3truecm}
\centerline{
    {Jaume Gomis,${}^{a}$}\footnote{jgomis@perimeterinstitute.ca}
    {Diego Rodr\'{\i}guez-G\'omez,${}^{b,c}$}\footnote{drodrigu@princeton.edu}
   {Mark Van Raamsdonk,${}^{d}$}\footnote{mav@phas.ubc.ca}
    {Herman Verlinde,${}^{b}$}\footnote{verlinde@princeton.edu}}
   \vspace{1.4cm}
\centerline{{\it ${}^a$Perimeter Institute for Theoretical Physics}}
\centerline{{\it Waterloo, Ontario N2L 2Y5, Canada}}
\vspace{.5cm}
\centerline{{\it ${}^b$Department of Physics,
Princeton University}}
\centerline{{\it Princeton, NJ 08544, USA}}
\vspace{.5cm}
\centerline{{\it ${}^c$Center for Research in String Theory, Queen Mary University of
    London}} \centerline{{\it Mile End Road, London, E1 4NS, UK}}
\vspace{.5cm}
\centerline{{\it ${}^d$Department of Physics and Astronomy,
University of British Columbia}}
\centerline{{\it 6224 Agricultural Road,
Vancouver, B.C., V6T 1Z1, Canada}}

\vspace*{1cm}
%%\maketitle
\end{center}

\centerline{\bf ABSTRACT}
\vspace*{10mm}
We test the proposals for the worldvolume theory of M2-branes by studying its maximally supersymmetric mass-deformation. We check the simplest prediction for the mass-deformed theory on $N$ M2-branes: that there should be a set of discrete vacua in one-to-one correspondence with partitions on $N$. For the mass-deformed Lorentzian three-algebra theory, we find only a single classical vacuum, casting doubt on its M2-brane interpretation. For the mass-deformed ABJM theory, we do find a discrete set of solutions, but these are more numerous than predicted. We discuss possible resolutions of this puzzling discrepancy.  We argue that the classical vacuum solutions of the mass-deformed ABJM theory display properties of fuzzy three-spheres, as expected from their gravitational dual   interpretation.

\end{titlepage}

\vskip 1cm

\addtolength{\baselineskip}{1.1mm}
\addtolength{\parskip}{.5mm}

\setcounter{equation}{0}
\section{Introduction}

The fact that M-theory has excitations corresponding to M2-branes and M5-branes is one of the most important clues to take into account in searching for a formulation of M-theory. Progress in the description of the long wavelength dynamics on multiple M2-branes should provide us with new important clues about the underlying degrees of freedom of M-theory.   Moreover, according to the AdS/CFT and Matrix theory holographic correspondences, the low-energy worldvolume theory on M2-branes gives a nonperturbative description of M-theory with $AdS_4 \times S^7$ boundary conditions, while     the   theory on  M2-branes compactified on a $T^2$ gives a nonperturbative formulation of the DLCQ of Type IIB string theory in flat spacetime.

The stimulating recent work of Bagger and Lambert \cite{Bagger:2006sk}\cite{Bagger:2007jr}\cite{Bagger:2007vi}  and Gustavsson \cite{Gustavsson:2007vu}\cite{Gustavsson:2008dy}
has resulted in the construction of novel three dimensional superconformal field theories, and in explicit proposals for how these field theories describe the effective low energy dynamics of M2-branes in various M-theory backgrounds.

Currently, there are two (seemingly) different proposals for the formulation of $N$ M2-branes in flat spacetime. Perhaps the most promising proposal \cite{Aharony:2008ug} -- henceforth called the ABJM proposal -- is based on an ${\cal N}=6$ $U(N)\times U(N)$ Chern-Simons gauge theory with matter interacting via a quartic superpotential. The theory with Chern-Simons   level $k=1$ has been proposed as the worldvolume description of $N$ M2-branes in flat spacetime. Despite the absence of an explicit  realization of the  required ${\cal N}=8$ supersymmetry (and $SO(8)$ symmetry), good evidence in favor of this proposal has been presented by ABJM \cite{Aharony:2008ug}.

An earlier proposal \cite{Gomis:2008uv}\cite{Benvenuti:2008bt}\cite{Ho:2008ei} is an ${\cal N}=8$ supersymmetric BF gauge theory with matter based on a Lorentzian three-algebra, henceforth called the BF membrane model.\footnote{A relation between the two proposals has been put forward in \cite{Honma:2008jd}.}  Even though this theory has ghosts, which nevertheless cannot propagate in loops, some evidence for this proposal has been gathered in \cite{Gomis:2008uv}\cite{Benvenuti:2008bt}\cite{Ho:2008ei}\cite{Banerjee:2008pd}\cite{Cecotti:2008qs}.\footnote{There is variant of this theory where a certain global symmetry is gauged in order to eliminate the ghosts \cite{Bandres:2008kj} \cite{Gomis:2008be}. This gauging does modify the theory and for a choice of prescription of how to perform the path integral one just recovers maximally supersymmetric Yang-Mills theory \cite{Bandres:2008kj}\cite{Gomis:2008be}\cite{Ezhuthachan:2008ch}, but in a reformulation where a formal $SO(8)$ superconformal symmetry is present. In the original version of \cite{Gomis:2008be}, it was suggested that full superconformal invariance could be maintained by integrating over a certain zero mode. We no longer feel that the justification for doing so is correct (see revised version of \cite{Gomis:2008be}).}  Despite this, many issues remain to be understood in this theory.

A challenge with verifying any proposal for the worldvolume theory of M2-branes is that the theory is strongly coupled, so generic calculations are not possible, even though one could in principle perform computer simulations by putting the theory on the lattice. In the case of ABJM, it is possible to modify the theory by changing the level from $k=1$ to large $k$ in order to get a weakly coupled theory, with $\lambda=N/k$ serving as an expansion parameter. The theory for general $k$ is proposed to describe M2-branes on an ${\cal N}=6$ preserving $\mathbb{R}^8/\mathbb{Z}_k$ orbifold, and some aspects of this prediction for large $k$ have been checked in \cite{Aharony:2008ug}\cite{Bhattacharya:2008bj}. However, a direct test of the theory for $k=1$ seems difficult. In particular, since $k$ is a discrete parameter, it is not even possible to compute supersymmetric indices at large $k$ to compare with $k=1$ predictions.

In this paper, we attempt to verify one nontrivial prediction for the $k=1$ theory. We recall that the M2-brane theory is known to possess a maximally supersymmetric mass-deformation preserving $SO(4) \times SO(4)$ global symmetry \cite{Bena:2004jw,Bena:2000zb}. For each of the two proposals, we attempt to verify that the mass-deformed theory has the expected  vacuum structure. In either the $T^2$ compactified theory or in the uncompactified case, the prediction is that the correct theory on $N$ M2-branes has a discrete set of vacua, in one-to-one correspondence with the partitions of $N$ \cite{Lin:2004nb,Mukhi:2002ck}. For the uncompactified theory, these vacua are dual to the bubbling geometries constructed by Lin, Lunin, and Maldacena in \cite{Lin:2004nb}. For the $T^2$ compactified theory, these vacua are interpreted as giant graviton states with zero energy and total longitudinal momentum $P^+ = N$ in the Hilbert space of Type IIB string theory in the maximally supersymmetric plane-wave background.

For the ABJM theory, we first need to derive the appropriate mass-deformation, which we do in section 2. Just as the undeformed theory does not possess the full $SO(8)$ invariance (and supersymmetry) in a manifest way, the mass deformed version cannot have more than an $SU(2) \times SU(2) \times U(1)$ subgroup of the $SO(4) \times SO(4)$ symmetry manifest. Nevertheless, using the AdS/CFT dictionary and known properties of the dual LLM geometries, we argue in section 2 for a particular form of the mass-deformation, corresponding to deforming the theory by  a dimension two relevant protected operator plus a dimension one operator that provides its supersymmetric completion. The theory we arrive at is precisely the same as a mass-deformation already written down in \cite{Hosomichi:2008jd}, which was shown in \cite{Hosomichi:2008jb}
to preserve as much supersymmetry as the original theory (that is ${\cal N}=6$ supersymmetry). Moreover, it was shown in \cite{Hosomichi:2008jd} that this mass-deformed   ABJM theory for    $SU(2) \times SU(2)$  gauge group
 reduces to the maximally supersymmetric and $SO(4)\times SO(4)$ invariant mass-deformed BF theory of \cite{Gomis:2008cv}\cite{Hosomichi:2008qk}.

Starting with this mass-deformed ABJM theory, we proceed in section 3 to consider its classical supersymmetric vacua. Here, we do find a non-trivial discrete set of vacua, but in this case, we have an embarrassment of riches: the vacua we find are more numerous than for the expected answer, the number of partitions of $N$. Specifically, we find that there are two irreducible solutions for each $N$ rather than the single one that would lead to the expected answer. In section 3.1, we discuss these results and suggest possible resolutions of the puzzling discrepancy. We note that the set of vacua we find are in one-to-one correspondence with a naive prediction for the number of vacua based on semiclassical expectations in the bulk dual, and we review the reasons why this naive prediction is not believed to be correct.

In section 3.2, we recall that the vacua of the mass-deformed theory should have an interpretation as fuzzy three-spheres, and give some evidence for this interpretation by constructing a ``radius-squared" operator which we show is proportional to the identity matrix for the irreducible solutions. Moreover, we show in section 3.3 that  the vacuum equations  in the massive theory allows us to also explicitly construct the sought-after ``fuzzy-funnels" \cite{Basu:2004ed} describing the M2-M5 brane intersection in the undeformed theory from the M2-brane worldvolume point of view. These classical solutions of the ABJM model describe how the M2-branes expand into a fuzzy $S^3$ near the core of the M5-branes. The identification of   solitons in the ABJM theory, and of vacua of the massive theory that have an M5-brane interpretation on fuzzy $S^3$'s opens up the possibility of studying the mysterious physics on multiple M5-branes by studying the physics of the gauge theory around these solutions.

The mass-deformation of the Bagger-Lambert theory with the right symmetries was written down in \cite{Gomis:2008cv}\cite{Hosomichi:2008qk}. In section 4, we find that this theory based on the Lorentzian three-algebra in \cite{Gomis:2008uv}\cite{Benvenuti:2008bt}\cite{Ho:2008ei} has a unique classical vacuum state. This result conflicts with M2-brane expectations, casting a shadow of doubt on the M2-brane interpretation of the BF membrane model. Quantum effects, however, may become important in analyzing the quantum vacua of  this theory (as in the ${\cal N}=1^*$ theory); these could certainly alter the classical result. As long as this discrepancy remains unresolved, however, it further undermines the viability of the BF membrane model. We close in section 5 with a discussion.

In    appendix $A$, we consider the vacua of the mass-deformed BF theory in the BRST-invariant treatment of \cite{Gomis:2008be}. In appendices $B,C$, we include various facts about the ABJM theory, including the supersymmetry transformations and the ${\cal N}=1$ and ${\cal N}=2$ superfield formulation of the theory. In appendix $D$
we study some other supersymmetric mass deformations of the ABJM theory, which should be dual to Type IIB plane-wave  backgrounds  with less supersymmetry.

\setcounter{equation}{0}

\section{The Mass Deformed ABJM Model}

Recently, ABJM have made an interesting proposal for the low energy description of $N$ coincident M2-branes in a
$\mathbb{C}^4/\mathbb{Z}_k$ orbifold. This theory is based on an ${\cal N}=6$ supersymmetric  $U(N)\times U(N)$ Chern-Simons gauge theory with levels $(k,-k)$ coupled to bifundamental matter, whose matter interactions are captured by a quartic superpotential.\footnote{When the gauge group is $SU(2)\times SU(2)$, the ABJM theory reduces to the original Bagger-Lambert theory in the Chern-Simons formulation of \cite{VanRaamsdonk:2008ft}.} It has been proposed that the theory for $k=1$ describes $N$  M2-branes  in flat spacetime and that the theory has enhanced ${\cal N}=8$ supersymmetry and $SO(8)$ R-symmetry.

The action for the ABJM theory may be written in a manifestly $SU(4) \times U(1)$ symmetric way as \cite{Benna:2008zy}
\begin{eqnarray}
S&=&\int d^3 x \left[ \frac{k}{4 \pi} \varepsilon^{\mu\nu\lambda} \mathrm{Tr} \left(
A_{\mu} \partial _{\nu} A_\lambda + \frac{2i}{3} A_{\mu} A_{\nu} A_{\lambda}
- \hat{A}_{\mu} \partial_{\nu} \hat{A}_{\lambda}
- \frac{2i}{3} \hat{A}_{\mu} \hat{A}_{\nu} \hat{A}_{\lambda} \right) \right. \cr
&&\left. - \mathrm{Tr} D_{\mu} C_I^{\dagger} D^{\mu} C^I
- i \mathrm{Tr} \; \psi^{I \dagger} \gamma^{\mu} D_{\mu} \psi_I \right] \cr
&+&\frac{4 \pi^2 }{3 k^2}
\mathrm{Tr} \left(
C^I C_I^\dagger C^J C_J^\dagger C^K C_K^\dagger
+ C_I^\dagger C^I C_J^\dagger C^J C_K^\dagger C^K \right. \cr
&&\left.
+4 C^I C_J^\dagger C^K C_I^\dagger C^J C_K^\dagger
-6 C^I C_J^\dagger C^J C_I^\dagger C^K C_K^\dagger \right)
\cr
&
+&\frac{2  \pi i }{k}
\mathrm{Tr} \left(
C_I^\dagger C^I \psi^{J \dagger} \psi_J
-\psi^{J \dagger} C^I C_I^\dagger \psi_J
-2 C_I^\dagger C^J \psi^{I \dagger} \psi_J
+2 \psi^{J \dagger} C^I C_J^\dagger \psi_I  \right. \cr
&&\left.
+\epsilon^{IJKL} C_I^\dagger \psi_J C_K^\dagger \psi_L
-\epsilon_{IJKL} C^I \psi^{J \dagger} C^K \psi^{L \dagger}
\right)\,.
\label{ABJM}
\end{eqnarray}

This action is invariant under a set of ${\cal N}=6$ supersymmetry transformations, which we write explicitly in appendix B. As we review in appendix C, this action may be written in terms of ${\cal N}=1$ superfields \cite{Hosomichi:2008jb}, or in terms of ${\cal N}=2$ superfields \cite{Aharony:2008ug,Benna:2008zy}. It may also be written as a three-algebra theory with complex structure constants \cite{Bagger:2008se}.

\subsection{Mass Deformed ABJM}

Assuming that ABJM gives a description of M2-branes in flat space, it should admit (at least for $k=1$)  a mass deformation which preserves maximal supersymmetry (albeit with a deformed supersymmetry algebra) and an $SO(4) \times SO(4) \times \mathbb{Z}_2$ subgroup of the original $SO(8)$ symmetry. Our first task will be to construct this deformation.

The deformed theory is expected to have a  discrete set of vacua labeled by partitions of $N$, where $N$ is the number of M2-branes. These vacua are dual to a discrete set of geometries (once flux quantization is taken into account) constructed in \cite{Lin:2004nb}.\footnote{These solutions were also studied in \cite{Bena:2004jw}.}
Since these geometries are all dual to vacua of the same field theory, they have the same asymptotic behavior (they are asymptotically $AdS_4 \times S^7$), and near the boundary differ from $AdS_4 \times S^7$ by exciting a non-normalizable mode of the four-form field strength of eleven dimensional supergravity. By the AdS/CFT dictionary
\cite{Maldacena:1997re}\cite{Gubser:1998bc}\cite{Witten:1998qj}, turning on a non-normalizable deformation of the supergravity four-form is dual (at the linearized level) to deforming the boundary field theory by a dimension two relevant operator. In addition, we expect mass terms for the scalars, so the deformed theory will contain dimension one operators as well. We will now try to explicitly write down these operators to derive the  maximally supersymmetric mass-deformation of the ABJM theory.

 Now, the undeformed ABJM theory has a manifest $SU(4) \times U(1)$ subgroup of the required $SO(8)$ R-symmetry group. The maximally supersymmetric mass-deformation breaks this $SO(8)$ to $SO(4) \times SO(4) \times \mathbb{Z}_2$. However, there are various inequivalent ways of embedding $SO(4) \times SO(4)$ into $SO(8)$ relative to a fixed embedding of $SU(4) \times U(1)$ in $SO(8)$. Thus, we expect that there should be a number of apparently inequivalent mass-deformations that are actually related by $SO(8)$ rotations.

To proceed, we will choose a particularly simple embedding as follows. Let $SO(8)$ act on a vector $v_i$. Then we define the preserved $SO(4) \times SO(4)$ subgroup to consist of $SO(8)$ transformations that do not mix the first four with the last four components of the vector $v_i$, while the $\mathbb{Z}_2$ to be the transformation that swap these components. Meanwhile, we can take the $SU(4)$ subgroup to consist of complex rotations on the four component complex vector $(v_1 + i v_2, v_3 + i v_4, v_5 + i v_6, v_7 + i v_8)$, and the remaining $U(1)$ subgroup to be the transformation that multiplies this complex vector by an overall phase. Now, the subgroup of $SO(8)$ common to  both  $SU(4) \times U(1)$ and  $SO(4) \times SO(4)$ is $SU(2) \times SU(2) \times U(1)$. This $SU(2) \times SU(2) \times U(1)$ is  generated by independent $SU(2)$ transformations on $(v_1 + i v_2, v_3 + i v_4)$ and $(v_5 + i v_6, v_7 + i v_8)$, and the overall phase rotation. These are clearly in $SO(4) \times SO(4)$, since they preserve the lengths of the two complex vectors. We conclude that the maximally supersymmetric mass-deformation should preserve a manifest $SU(2) \times SU(2) \times U(1) \times \mathbb{Z}_2$ symmetry.
%We can take these two $SU(2)$s to be the ones that rotate  $A_\alpha$ and $B_{\dot{\alpha}}$ in ABJM, where $\alpha,\dot{\alpha}$ is a doublet index for each of the $SU(2)$'s.

Since the mass-deformation should leave us with no flat  directions, we expect that it should give masses to all the scalar fields. In order to maintain the desired  $SU(2) \times SU(2) \times U(1) \times \mathbb{Z}_2$ invariance, the only possibility is that all the scalars have equal masses. Therefore, we have to add the  following mass term\footnote{One might be concerned that this operator is not protected. However, this operator  arises at second order in the deformation parameter $\mu$, and we will see that it can be thought of as the supersymmetric completion of the leading order deformation (a dimension two operator proportional to $\mu$),  which we will demand is a protected operator. We thank Shiraz Minwalla for a discussion on this point.}
\be
\label{ABmass}
{\cal L}_{mass} = \mu^2 \tr(Q^\alpha Q^\dagger_\alpha + R^\alpha R^\dagger_\alpha)  = \mu^2 \tr(C^I C^\dagger_I) \; ,
\ee
where we have used that
\be
C^I\equiv(Q^1,Q^2,R^1,R^2).
\ee
In addition, we should also add a  fermion mass terms and a quartic term in the potential which together deform the   theory by a dimension two operator. We can think of this combination as the leading order deformation since it is linear in
the deformation parameter $\mu$.

There are various ways to pin down the form of this dimension two operator that we must deform by. We know that deforming with  this operator is related holographically to turning on the non-normalizable mode four-form field strength of eleven dimensional supergravity, since this deformation is present in the dual LLM geometries. This operator should be a protected operator, but the only $SO(8)$ multiplets of protected operators of dimension two that are worldvolume scalars are the chiral primary operators in the four-index symmetric traceless representation of $SO(8)$ (the ${\bf 294}_v$), and operators in the self-dual antisymmetric tensor representation of $SO(8)$ (the ${\bf 35}_c$), obtained by the action of two supersymmetries on the dimension one chiral primary operator in the symmetric traceless two-index representation of $SO(8)$ (the ${\bf 35}_v$). Since the operator we are interested in deforming by is dual to a four-form field strength in the bulk whose non-zero components have all legs transverse to the $SO(8)$, it must be some component of the latter set of operators.\footnote{
Another way of thinking about the mass-deformation is as follows. The torus-compactified M2-brane theory should provide the Matrix theory description of Type IIB string theory in flat space
\cite{Sethi:1997sw}\cite{Banks:1996my}, while the maximally supersymmetric mass-deformed theory on $T^2$ should give the Matrix theory description of Type IIB string theory on the maximally supersymmetric plane-wave. The latter is obtained by turning on $h_{++} = \mu^2 x^I x^I$ and $F_{+1234}=F_{+5678}=\mu$. As in the BFSS matrix model, we expect that for every (off-shell) linear deformation of the bulk geometry, there should be a corresponding operator that we can add to the Lagrangian of the Matrix theory description. Such a correspondence was worked out for BFSS matrix theory
\cite{Banks:1996vh} in \cite{Taylor:1998tv}. In that case, turning on $h_{++} \propto M_{AB} x^A x^B$ (as one does to obtain the maximally supersymmetric eleven-dimensional plane wave) corresponds to adding bosonic mass terms to the BFSS Lagrangian, while turning on constant eleven-dimensional supergravity flux $F_{+123}$ corresponds to adding fermion mass terms plus bosonic interactions. In our case, we similarly expect that the mass terms (\ref{ABmass}) correspond to the $h_{++}$ field in the Type  IIB plane-wave, while the quartic interaction term that we would like to construct (and the fermion mass term) is the operator associated with the supergravity field $F_{+1234}=F_{+5678}=\mu$. For a moment, let us consider the operator ${\cal O}^{IJKL}$ corresponding to turning on a more general five-form flux $F_{+IJKL}$ in the Type IIB theory. The self-duality of the five-form implies that
\[
F_{+IJKL} = \epsilon_{IJKL}{}^{MNPQ}F_{+MNPQ}
\]
Therefore, this field is in the ${\bf 35}_c$ representation of $SO(8)$, so we expect that the corresponding operator will also be in this representation, as we argued earlier.}

Given the set of operators filling out the ${\bf 35}_c$ representation of $SO(8)$, it turns out that there is a unique linear combination that preserves $SU(2) \times SU(2) \times U(1)$.\footnote{With our conventions above, the desired combination is ${\cal O}^{1234} + {\cal O}^{5678} $.} To see this, note that under $SU(4) \times U(1)$, we have that
\[
{\bf 35}_c \to {\bf 10}_{-2} + {\bf \bar{10}}_{2} + {\bf 15}_0 \; ,
\]
and under the decomposition of $SU(4)$ to $SU(2) \times SU(2)$, we have\footnote{The ${\bf 10}$ and  ${\bf \bar{10}}$ representations of $SU(4)$ have no singlets under $SU(2) \times SU(2)$.}
\[
{\bf 15}_0 = ({\bf 2},{\bf 2})+({\bf 3},{\bf 1})+({\bf 1},{\bf 1})+({\bf 1},{\bf 3})+({\bf 2},{\bf 2}) \; ,
\]
so we want to turn on the singlet operator in this decomposition.

More explicitly, if we can denote the set of operators in the ${\bf 15}_0$ representation of $SU(4) \times U(1)$ by
\[
{\cal O}^I_J,
\]
 then the $SU(2) \times SU(2)$ singlet we are looking for is
\[
{\cal O}^\alpha_\alpha = - {\cal O}^{\dot{\beta}}_{\dot{\beta}}.
\]
To find the form of the operators ${\cal O}^I_J$, we can use the fact that they may be obtained by acting with  supersymmetry transformations on the dimension one chiral primary operators, with transform in the ${\bf 35}_v$ representation of $SO(8)$. Schematically, the  supersymmetry transformations act as
\beas
\ba{ccccc}
{\rm dim 1} && {\rm dim 3/2} && {\rm dim 2} \cr
{\bf 35}_v &\rightarrow_Q &{\bf 56}_s &\rightarrow_Q &{\bf 28}_v + {\bf 35}_c
\ea
\eeas
The ${\bf 28}_v$ is straightforward to distinguish from the ${\bf 35}_c$, since the latter is a worldvolume scalar, while the former is a worldvolume vector. In particular, the ${\bf 35}_c$  comes from acting with a combination of two supersymmetry generators whose ${\bf 8}_s$ spinor indices of $SO(8)$ are symmetrized.

Under $SU(4) \times U(1)$, the dimension one chiral primary operators decompose as follows
\[
{\bf 35}_v \to {\bf 10}_2 + {\bf 15}_0 + \bar{{\bf 10}}_{-2} \; .
\]
The operators in the ${\bf 15}_0$ representation may be written as \cite{Aharony:2008ug}
\be
\label{primary}
{\cal P}^I_J = \tr(C^I C^\dagger_J) - {1 \over 4} \delta^I_J \tr(C^K C^\dagger_K) \; .
\ee
The remaining operators in the multiplet are charged under the $U(1)$, and gauge invariance requires the insertion of undetectable  Wilson lines running from the location of the operator off to infinity, as explained in  \cite{Aharony:2008ug}. These exist only for $k=1$. In a similar way, the operators in the ${\bf 35}_c$ that are charged under $U(1)$ also require Wilson lines, but we are interested in the uncharged operators ${\cal O}^I_J$ and we should be able to write these as conventional operators.

We then expect that they can be obtained by acting with the conventional supersymmetry transformations (in the ${\bf 6}$ of $SU(4)$) on the conventional components of the chiral primary operator (\ref{primary}). To summarize, we want to act on the operator (\ref{primary}) in the ${\bf 15}_0$ representation of $SU(4) \times U(1)$ with two supersymmetry operators whose group theory indices are symmetrized to obtain a descendent operator which is also in the ${\bf 15}_0$ representation of $SU(4) \times U(1)$. This uniquely specifies the operator as
\[
{\cal O}^I_J \propto ( Q_{JL} Q^{IN} + Q^{IN} Q_{JL}- {1 \over 4} \delta^I_J(Q_{PL} Q^{PN} + Q^{PN} Q_{PL})) {\cal P}^L_N,
\]
where the spinor indices on the $Q$s are contracted. Using the supersymmetry transformations in the appendix, this gives:
\[
{\cal O}^I_J = {16 \pi \over k} \tr(C^K C^\dagger_{[K} C^I C^\dagger_{J]})- {1 \over 4} \delta^I_J \tr(C^K C^\dagger_{[K} C^L C^\dagger_{L]}) + 2\tr(\psi^{\dagger I} \psi_J - {1 \over 4}\delta^I_J \psi^{\dagger K} \psi_K)\,.
\]
The operator we wish to turn on is the $SU(2) \times SU(2)$ singlet component of this, proportional to
\[
{\cal O}^\alpha_\alpha = -{\cal O}^{\dot{\alpha}}_{\dot{\alpha}} = {8 \pi \over k}\tr(Q^\alpha Q^\dagger_{[\alpha}Q^\beta Q^\dagger_{\beta]} - R^\alpha R^\dagger_{[\alpha}R^\beta R^\dagger_{\beta]}) + \tr(\psi^{\dagger \alpha} \psi_\alpha - \chi^{\dagger \dot{\alpha}} \chi_{\dot{\alpha}})\,,
\]
where we have written $C^I = (Q^\alpha, R^\alpha)$. The coefficient of this operator in the deformed Lagrangian is proportional to $\mu$ and can be fixed by demanding that the full bosonic potential is a sum of squares as we expect from  supersymmetry. There is indeed such a choice, which gives
\bea
V = |M^\alpha|^2 + |N^\beta|^2,
\label{potenini}
\eea
where
\beas
M^\alpha &=&  \mu Q^\alpha + {2 \pi\over k }(2Q^{[\alpha} Q^\dagger_\beta Q^{\beta]} + R^{\beta} R^\dagger_\beta Q^\alpha- Q^\alpha R^\dagger_\beta R^{\beta} + 2 Q^\beta R^\dagger_\beta R^\alpha - 2R^\alpha R^\dagger_\beta  Q^\beta ) \cr
N^\alpha &=&  - \mu R^\alpha + {2 \pi \over k}(2R^{[\alpha} R^\dagger_\beta R^{\beta]} + Q^{\beta} Q^\dagger_\beta R^\alpha- R^\alpha Q^\dagger_\beta Q^{\beta} + 2 R^\beta Q^\dagger_\beta Q^\alpha - 2Q^\alpha Q^\dagger_\beta  R^\beta ).
\eeas

It turns out that this mass deformation is exactly the one considered in using the ${\cal N}=1$ superfield formalism in \cite{Hosomichi:2008jd}, as we show in appendix C. In that language, this deformation is captured by a deformation of the superpotential, and the expressions ${\cal M}$ and ${\cal N}$ above appear as
\beas
M^\alpha &=& {\partial W \over \partial Q^\dagger_\alpha} \cr
N^\alpha &=& {\partial W \over \partial R^\dagger_\alpha},
\eeas
where $W=W_{ABJM}+W_{def}$ (see appendix $C$ for their explicit expressions).

\section{Vacua of the mass-deformed ABJM  theory}

Now, consider the vacuum equations in the mass-deformed ABJM theory. For the potential in (\ref{potenini}) to vanish, we require
\beas
\mu Q^\alpha + {2 \pi  \over k}(2Q^{[\alpha} Q^\dagger_\beta Q^{\beta]} + R^{\beta} R^\dagger_\beta Q^\alpha- Q^\alpha R^\dagger_\beta R^{\beta} + 2 Q^\beta R^\dagger_\beta R^\alpha - 2R^\alpha R^\dagger_\beta  Q^\beta ) &=& 0 \cr
- \mu R^\alpha + {2 \pi \over k}(2R^{[\alpha} R^\dagger_\beta R^{\beta]} + Q^{\beta} Q^\dagger_\beta R^\alpha- R^\alpha Q^\dagger_\beta Q^{\beta} + 2 R^\beta Q^\dagger_\beta Q^\alpha - 2Q^\alpha Q^\dagger_\beta  R^\beta ) &=& 0\,.
\eeas
Note that for block diagonal matrices,  these equations split into separate equations of the same form for each block. Thus, we can build up solutions for our $N \times N$ matrices from solutions for smaller matrices. The expected set of discrete solutions for the mass-deformed M2-brane theory, in one-to-one correspondence with the partitions of $N$, will then result if there is a single irreducible solution for each $N$.

While these equations look somewhat complicated in general, it is straightforward to find nontrivial solutions if we assume that either $R=0$ or $Q=0$.\footnote{We believe that these are the only solutions, but we have not proven it.} In these cases, we need only solve
\be
\label{qeq}
Q^\alpha +  Q^{\alpha} Q^\dagger_\beta Q^{\beta} - Q^{\beta} Q^\dagger_\beta Q^{\alpha} = 0
\ee
or
\be
\label{req}
R^\alpha -  R^{\alpha} R^\dagger_\beta R^{\beta} + R^{\beta} R^\dagger_\beta R^{\alpha}= 0
\ee
respectively, where we have set $\mu k /(2\pi) =1$ for notational convenience. Here, the first equation may be obtained from the second by Hermitian conjugation and the map $R^\dagger_\alpha \to Q^\alpha$, so solutions of (\ref{qeq}) and (\ref{req}) are in one-to-one correspondence. More explicitly, we can see that making the substitution $R_\alpha = A_\alpha$ in (\ref{req}) or $Q_\alpha = A^\dagger_\alpha$ in the adjoint of (\ref{qeq}) give the equations
\bea
\label{eqns}
A_1 + A_2 A_2^\dagger A_1 - A_1 A_2^\dagger A_2 &=& 0 \cr
A_2 + A_1 A_1^\dagger A_2 - A_2 A_1^\dagger A_1 &=& 0\,.
\eea
To solve these, we note first that by a $U(N)\times U(N)$ gauge transformation
\[
A_1 \to U A_1 V^{-1} \;
\]
we can bring $A_1$ to a diagonal matrix with nonnegative real entries, which are strictly increasing
\[
A_1 = {\rm diag}(x_1, \dots, x_N) \qquad \qquad 0 \le x_1 \le \cdots \le x_N\,.
\]
Now, the second equation gives
\[
(A_2)_{ab}(1 + x_a^2 - x_b^2) = 0\,.
\]
From this, it follows that $(A_2)_{ab}$ can be nonzero if and only if
\[
x_b^2 = 1 + x_a^2 \; .
\]
Since we have arranged that the $x$'s are ordered, this also implies that $a<b$.

Now, let us assume that the matrices $A_1$ and $A_2$ form an irreducible solution of our equations. Consider first the case where all $x_i$'s are different. Then it must be that
\be
\label{seq}
x_1^2 = x_2^2 - 1 = x_3^2 - 2 = \dots \; .
\ee
Otherwise, the sequence $(x_{k_1},x_{k_2}, x_{k_3}, \dots)$ of $x$'s such that the square of each is one greater than the square of the previous would be missing some set $(x_{l_1}, \dots, x_{l_m})$ of $x$'s. In this case, all matrix elements $(A_2)_{k_i l_j}$ would have to vanish, contradicting our irreducibility assumption.

Now consider the case where some of the $x$'s are the same. If we have $x_k = x_{k+1} = \cdots = x_{k+m}$ and $x_l = x_{l+1} = \cdots = x_{l+n}$, and $x_l^2 = x_k^2 - 1$ then in principle, we could have any of the matrix elements $(A_2)_{k+i,l+j}$ nonzero. But in this case, we have residual gauge symmetry that can be used to make all elements of this $(m+1) \times (n+1)$ matrix vanish except the ``diagonal'' $i=j$ elements. It is then straightforward to see that the matrix is again reducible.

Thus, in the irreducible case, we may assume that the elements of $A_1$ satisfy (\ref{seq}), and it follows that the only nonzero elements of $A_2$ are $(A_2)_{i \;  i+1} = a_i$. We can use gauge transformation for which $U=V$ is diagonal to make all of these elements real and positive. Using the first equation in (\ref{eqns}), we find that
\[
x_i (1 + a_i^2 - a_{i-1}^2) = 0
\]
These imply a unique solution
\[
x_n = \sqrt{n-1} \qquad \qquad a_{n} = \sqrt{N-n} \; ,
\]
or more explicitly,
\be
A_1 = \left( \ba{ccccc} 0 &&&& \cr &1 &&& \cr && \ddots &&\cr &&& \sqrt{N-2} &\cr &&&& \sqrt{N-1} \ea \right) \qquad \qquad A_2 = \left( \ba{ccccc} 0 & \sqrt{N-1} &&& \cr &0 & \ddots&& \cr && \ddots &\sqrt{2} &\cr &&&0 & 1 \cr &&&& 0  \ea \right)
\ee

Thus, we have a single irreducible solution $A_N$ to the equation (\ref{eqns}) for each integer $N$.
But since every solution to $(\ref{eqns})$ gives a solution to (\ref{qeq}) (equivalent to the full set of equations in the case $R=0$) and to (\ref{req}) (equivalent to the full set of equations for $Q=0$), we have actually found two inequivalent irreducible solutions for each $N>1$, namely $(Q,R) = (0,A_N)$ and $(Q,R) = (A_N^\dagger,0)$, together with the trivial irreducible solution $(Q,R)=(0,0)$ for $N=1$.
Correspondingly, we have a set of reducible solutions to the full equations of the form
\be
Q = \left( \ba{cccccc} A_{N_1}^\dagger &&&&& \cr &\ddots&&&& \cr && A_{N_k}^\dagger&&& \cr &&& 0 && \cr &&&& \ddots & \cr &&&&& 0 \ea \right) \qquad \qquad R = \left( \ba{cccccc} 0 &&&&& \cr &\ddots&&&& \cr && 0 &&& \cr &&& A_{N_{k+1}} && \cr &&&& \ddots & \cr &&&&& A_{N_m} \ea \right)
\ee
where we allow $N_l=1$ only for $l>k$. These solutions are clearly more numerous than partitions of $N$, so there is a discrepancy between the counting of classical vacua and the expected number of vacua for the theory. We discuss this in detail presently.

\subsection{Possible resolutions}

Ironically, the set of solutions we have found for the deformed ABJM  theory corresponds well with a naive picture of how many solutions there should be. To see this, let us discuss specifically the case where the theory is defined on $T^2$ such that the interpretation is the DLCQ of Type IIB string theory on the maximally supersymmetric plane-wave. In this case, the vacua are interpreted as giant gravitons in the plane-wave background. There are two types of giant gravitons, those that expand in the $1234$ directions and those that expand in the $5678$ directions. For the uncompactified plane-wave, the general ground states with fixed $P^+$ contain concentric giant gravitons  with various radii in the $1234$ directions and concentric giant gravitons with various radii in the $5678$ directions. In the DLCQ of the Type IIB theory, a given giant graviton will carry some integer number of units of $P^+$, and the total number of $P^+$ units must equal $N$. Thus, we might conclude that the set of vacua should be in correspondence with the number of ways of distributing the total momentum $N$ discretely between giant gravitons, each of which lives either in the $1234$ directions or in the $5678$ directions. This matches exactly with the counting that we have found, assuming that the $1234$ giant graviton with 1 unit of $P^+$ is the same as the $5678$ giant graviton with 1 unit of $P^+$ (we could call this an ordinary graviton).

However, the analysis in the previous paragraph is probably too naive since it relies on the classical picture of giant gravitons which we can trust only in the large $N$ limit. To explain this more precisely, we note that one could have made a similar argument for M-theory on the plane-wave, whose ground states for fixed $P^+$ include concentric membranes expanded in three directions together with concentric fivebranes expanded in six transverse directions. In this case, for finite $N$, the field theory description (the plane-wave matrix model \cite{Berenstein:2002jq}\cite{Dasgupta:2002hx}) shows that the ground states are labeled simply by partitions of $N$, rather than labeled partitions. The explanation is that at finite $N$ the distinction between membrane states and fivebrane states is ambiguous. For example, the state corresponding to $k$ concentric fivebranes each carrying $m$ units of momentum is the same as the state with $m$ membranes carrying $k$ units of momentum. It is only in the large $N=mk$ limit, holding either $k$ fixed or $m$ fixed that one is unambiguously describing fivebranes or membranes respectively. It is reasonable to expect that the present case works in the same way, so the prediction that the massive M2-brane vacua should be in correspondence with partitions of $N$, based on the LLM analysis and the quiver gauge theory description of the DLCQ of Type IIB on a plane wave, seems robust.

Assuming that the ABJM theory does provide a correct description of the worldvolume physics of M2-branes and that our classical analysis is correct, there are various possible conclusions here:
\begin{itemize}
\item

The prediction that the vacua should be labeled by partitions of N is
incorrect. This comes independently from construction of the dual geometries in \cite{Lin:2004nb} and from the BMN quiver gauge-theory description of the DLCQ of the Type IIB plane-wave \cite{Mukhi:2002ck}, and as we have argued above, agrees with the plane-wave matrix model in which we have similar physics (thought in that case, there is no symmetry between the two types of giant gravitons).\footnote{We should note that there is actually a third proposal \cite{SheikhJabbari:2004ik} for the DLCQ of the Type IIB plane-wave (see also
\cite{Lozano:2006jr}), in which the counting of vacua is exactly the same as we have found in ABJM.}

\item

The mass-deformation we have considered is not the maximally supersymmetric one. Since the deformation we consider reduces to the maximally SUSY deformation in at least one special case, the only possibility is that we have missed some terms in the deformation that vanish in the case of $SU(2) \times SU(2)$ gauge group.  An interesting possibility is that the full operator ${\cal O}^I_J$ in the ${\bf 15}_0$ representation of $SU(4) \times U(1)$ that we need to add also includes non-conventional terms with Wilson lines. However, adding these extra terms would have to preserve all of the supersymmetry already present. The presence of such terms would mean that the mass-deformation does not make sense for arbitrarily large values of $k$, since Wilson lines are no longer local for large enough $k$. This possibility would be ruled out if there exist LLM-type geometries arising as deformations of $AdS_4 \times S^7/\mathbb{Z}_k$, but we do not know whether this is the case.

\item

We are analyzing the vacua too naively. Since our analysis has been completely classical, we may worry that the results are not robust for the quantum theory. In the ${\cal N}=1^*$ theory considered by Polchinski and Strassler \cite{Polchinski:2000uf}, the classical counting of vacua does not give the whole story, since each classical vacuum splits into some number of quantum vacua. In our case, we require quantum effects to eliminate some of the vacua. This seems unlikely with this large amount of supersymmetry, but we cannot rule it out. It seems quite useful to understand what the prediction is for the number of vacua in the theory at large $k$, since here the theory is weakly coupled and one should have control over any possible quantum corrections.

\end{itemize}

We should note that the story may be even more complicated for the case of the mass-deformed theory on $T^2$. In general, the classical vacua preserve some of the gauge symmetry. If we expand the theory about one of these vacua in the large mass limit,
then it would appear that the low-energy physics is just pure Chern-Simons theory on $T^2$ for the preserved gauge-group.

Now, pure $U(N)$ CS theory on $T^2$ has a Hilbert space of dimension
\[
\left( \ba{c} N+k \cr k \ea \right)
\]
(the states correspond to the integrable representations of the affine Lie algebra corresponding to the gauge group $G$ at level k). If this degeneracy appears also in our theory, each classical vacua would have an additional quantum degeneracy depending on the level and the preserved gauge group, similar to the  $N=1^*$ theory. However, the formula above takes into account the quantum level shift, and this is known to be regulator dependent. In our case, the heavy massive matter fields can be thought of as a supersymmetric regulator for the Chern-Simons theory. It seems plausible that there would be no level shift in this case, and that the dimension of the Hilbert space would be one at least in the $k=1$ case.

\subsection{Interpretation as a three-sphere}

Vacuum states of the mass-deformed M2-brane theory have a natural interpretation as fuzzy $S^3$s. The dual geometries can be thought of as the near horizon geometries corresponding to dielectric M2-branes puffing into M5-branes in which three of the directions form a fuzzy $S^3$ \cite{Bena:2000zb}, similar to Polchinski-Strassler \cite{Polchinski:2000uf}. The LLM supergravity solutions \cite{Lin:2004nb}\cite{Bena:2004jw} capture the complete backreaction of this  configuration of dielectric M5-branes when the number of fivebranes is sufficiently large. Alternately, in the $T^2$ compactified case, these vacua correspond to D3-brane giant gravitons. In the large $N$ limit, these have a classical spherical geometry, but for finite $N$ (in the DLCQ theory) we expect that these are fuzzy spheres.

We will now see that the classical solutions we have found have features consistent with their interpretation as spherical brane configurations. To see this, we first recall that in the similar situation of Dp-branes expanded into D(p+2)-branes, a way to see that the configuration $X^i \propto J^i$ corresponds to a fuzzy two-sphere geometry is to notice that
\[
X_1^2 + X_2^2 + X_3^2 = R^2 \identity_{N \times N}
\]
We can interpret the left side as a radius-squared operator, and we see that all eigenvalues are the same.

Since the matter fields in ABJM are complex matrices in the bifundamental representation, it is natural to define the radius squared operators
\be
L_1^2=Q_\alpha Q^\dagger_{\alpha}\qquad L_2^2=R_\alpha R^\dagger_{\alpha},
\ee
where the first gives the radius of the $S^3$ in the 1234 directions and the second gives the radius of the $S^3$ in the 5678 directions. A straightforward computation shows that for our $Q$ vacua,
\be
L_1^2 = (N-1) \cdot \identity
\ee
while for the $R$ vacua $L_2^2=(N-1)$,
which can be interpreted as an $S^3$ with radius $\sqrt{N-1}$.

The interpretation that the classical vacua correspond to M2-branes expanded into M5-branes opens the prospect of studying the physics of multiple coincident M5-branes by studying the physics of the mass-deformed M2-brane theory around one of these vacua.

\subsection{An aside: fuzzy funnels}

One of the precursors that led the construction of the Bagger-Lambert theory is the study of the description of the M2-M5 brane intersection
\begin{equation}
\begin{array}{cccccccccccc}
   & 0 & 1 & 2 & 3 & 4& 5 & 6 & 7 & 8 & 9 & 10\\
\mbox{M2:} & \times & \times & \times & & & & &  &  & & \\
\mbox{M5:} & \times & \times &  & \times & \times  & \times
& \times &   &  &   \\
\end{array}
\label{confi}
 \end{equation}
from the point of view of the worldvolume of the M2-branes  \cite{Basu:2004ed} (for another precursor see
\cite{Schwarz:2004yj}). The equation describing this brane configuration from the point of view of the M2-branes is the analog of the ``fuzzy funnel" equation describing the D1-D3 intersection, which describes the   D1-branes blowing up into a fuzzy $S^2$ near the D3-brane core \cite{Constable:1999ac}.

Given this M2-M5 brane configuration, the M2-brane is expected to expand into a fuzzy $S^3$ near the M5-brane core. So far, we have analyzed vacua of the mass-deformed ABJM, which also have an interpretation as fuzzy $S^3$'s. It is therefore not a coincidence that the ``fuzzy funnel" solutions  of the undeformed theory are very closely related to the vacua of the mass-deformed ABJM theory studied in this paper.

The ``fuzzy funnel" equation describing the M2-M5 brane intersection above corresponds from the point of view of the field theory on the M2-branes to a BPS domain wall.  In the ${\cal N}=1$ superfield formulation, the superpotential of the theory is given by (\ref{superone}). The
``fuzzy funnel" equation corresponding to the brane configuration (\ref{confi}) is therefore given by
\bea
{dQ^\alpha\over dx_2}={\partial W\over \partial Q_{\alpha}^\dagger}={\pi k\over 2}\left(Q^\alpha Q_\beta^\dagger Q^\beta-Q^\beta Q_\beta^\dagger Q^\alpha\right),
\label{funnel}
\eea
where since the  M5-brane sits at $x_7=x_8=x_9=x_{10}=0$ we have set $R^\alpha=0$. There is an analogous equation  to (\ref{funnel}) describing an M5-brane sitting at $x_3=x_4=x_5=x_{6}=0$ where $Q^\alpha\leftrightarrow R^\alpha$.

The spatial profile of the ``fuzzy funnel"  can be trivially solved by $Q^{\alpha}\rightarrow {1\over \sqrt{x_2}}Q^{\alpha}$, and what the are left with is precisely the equations (\ref{qeq}) for the vacua of the mass-deformed ABJM model. We can then use our classification of these vacua to characterize the solutions describing ``fuzzy funnels" in the M2-brane theory.\footnote{The fuzzy funnel solution for $N=2$ has recently been considered in \cite{Terashima:2008sy}}

\section{Mass Deformed BF Membrane Model}

 We now turn to a discussion the mass-deformed BF membrane model. The starting point is the theory written down in \cite{Gomis:2008uv}\cite{Benvenuti:2008bt}\cite{Ho:2008ei} based on the Bagger-Lambert-Gustavsson construction with  a Lorentzian three algebra. This theory has ${\cal N}=8$ supersymmetry, in fact it is invariant under
   the $Osp(8|4)$ supergroup. This theory was conjectured to provide a worldvolume  description on $N$ coincident M2-branes in flat spacetime.

\subsection{BF Membrane Model}

  The action for the corresponding theory is given by
\begin{eqnarray}
\cal{L}\is -\frac{1}{2}{\rm Tr}\Big((D_{\mu}X^I - B_\mu X^I_+)^2 \Big) + \partial_{\mu} X^I_+ (\partial_{\mu} X_-^I - {\rm Tr}(B_\mu X^I)) + \frac{i}{2}{\rm Tr}\Big(\bar{\Psi}\Gamma^{\mu}(D_{\mu}\Psi- B_\mu \Psi_+)\Big)\nonumber \\[1mm] &&- \frac{i}{2}\bar{\Psi}_+\Gamma^{\mu}(\partial_{\mu}\Psi_- - {\rm Tr}( B_\mu \Psi)) -\frac{i}{2}\bar{\Psi}_-\Gamma^{\mu}\partial_{\mu}\Psi_+ + \epsilon^{\mu\nu\lambda} {\rm Tr}\Big( B_{\lambda} (\partial_{\mu} A_{\nu} - [A_{\mu}, A_{\nu}]) \Big) \nonumber\\[1mm]
&& - \frac{1}{12} {\rm Tr}\Big(X_+^I [ X^J, X^K] + X^J_+ [ X^K , X^I ] + X_+^K [ X^I , X^J ]\Big)^2 \label{action}
\\[1mm]
&& +\frac{i}{2}{\rm Tr}\Big(\bar{\Psi}\Gamma_{IJ}X_ +^I[X^J,\Psi]\Big)+\frac{i}{4}{\rm Tr}\Big(\bar{\Psi}\Gamma_{IJ}[X^I,X^J]\Psi_+\Big)-\frac{i}{4}{\rm Tr}\Big(\bar{\Psi}_+\Gamma_{IJ}[X^I,X^J]\Psi\Big)\;.
\nonumber
\end{eqnarray}
Here $D_\nu = \partial_\nu -2 [A_\nu, \cdot\, ]$,
with $A_\mu$  a gauge field for the compact gauge group $G$, which we will take to be $U(N)$.
The fields  $X^I$, $\Psi$, and $B_\mu$ transform in the adjoint representation for this gauge symmetry,
while the fields $X^I_+$,$X^I_-$,$\Psi_+$, and $\Psi_-$ are singlets. The above
action does not contain a standard Yang-Mills kinetic term for the gauge boson $A_\mu$, but rather
a term of the  $B \wedge F$ form, which underlies the symmetry structure of  (\ref{action}). Via the presence of this additional one-form field $B$,
the theory has an extra non-compact gauge symmetry, under which the fields transform infinitesimally as
\bea
\label{Bgauge}
\deltaa B_\mu \is D_\mu \zeta\, ; \qquad \qquad
\deltaa X^I = \, \zeta X^I_+ \, ; \qquad \qquad
\deltaa X^I_-  =\, \tr(\zeta X^I) \, ; \nonumber\\[1mm]
\deltaa \Psi \is \zeta \Psi_+ \, ; \qquad \qquad \deltaa \Psi_- = \tr(\zeta \Psi) \, .
\eea
The non-compact and compact symmetry  together combine into a gauge invariance under the
(2 dim $G$)-dimensional gauge group given by the Inonu-Wigner contraction of $G \otimes G$, which corresponds to the semidirect sum of the of translation algebra with the Lie algebra $G$, where $G$ acts on the $\hbox{dim} \; G$ translation generators in the obvious way.

\subsection{Mass Deformation}

 There exists a one parameter deformation of the BF membrane model which is maximally supersymmetric and $SO(4)\times SO(4)$ invariant \cite{Gomis:2008cv}\cite{Hosomichi:2008qk}.  The mass-deformed theory adapted to   Lorentzian three-algebras  is obtained by adding the following terms to the undeformed action
\beas
{\cal L}_{mass} &=& \mu^2 X^I_-X^I_+-{\mu^2\over 2}{\rm Tr}(X^IX^I)-{i}\mu \bar{\Psi}_+\Gamma_{3456}\Psi_-
+{i\over 2}\mu {\rm Tr}(\bar{\Psi}\Gamma_{3456}\Psi)\\
{\cal L}_{flux}&=&-{2\over 3}\mu\hskip+1pt \varepsilon^{ABCD}X_+^A{\rm Tr}\Big([X^B,X^C]X^D\Big)-{2\over 3}\mu\hskip+1pt \varepsilon^{A'B'C'D'}X_+^{A'}{\rm Tr}\left([X^{B'},X^{C'}]X^{D'} \right)\,.
\eeas
The eight scalars split into two groups  $X^I=(X^A,X^{A'})$, acted on by $SO(4)\times SO(4)$.
This maximally supersymmetric deformation gives mass to all the scalar and fermions in the theory as well as turning on
a quartic potential for the scalars, induced by a background four-form flux.

Using the deformed supersymmetry transformations of the full theory,\footnote{The supersymmetry algebra appeared in the general Nahm classification \cite{Nahm:1977tg}, see e.g. \cite{Lin:2005nh} for the explicit form of the algebra.}
it was shown in \cite{Gomis:2008cv}\cite{Hosomichi:2008qk} that the
set of maximally supersymmetric vacua are given by homogenous field configurations solving the following equations
\bea
[X^A,X^B,X^C]=-\mu\hskip+1pt \epsilon^{ABCD}X^D,\qquad X^{A'}=0
\label{vacua}
\eea
and  alternatively
\bea
[X^{A'},X^{B'},X^{C'}]=-\mu\hskip+1pt \epsilon^{A'B'C'D'}X^{D'},\qquad
X^{A}=0.
\eea
Here we want to solve these equations for the theory based on the Lorentzian algebra of \cite{Gomis:2008uv}\cite{Benvenuti:2008bt}\cite{Ho:2008ei}. The equations (\ref{vacua})  are then given by
\bea
\tr([X^A, X^B] X^C) &=& -\mu\epsilon^{ABCD} X^D_- \cr
3 X^{[A}_+ [X^B,X^{C]}] &=& -\mu \epsilon^{ABCD} X^D \cr
0 &=& -\mu \epsilon^{ABCD} X^D_+,
\label{lorenzo}
\eea
with an identical set of equations with $X^A\rightarrow X^{A'}$.
The last equation in (\ref{lorenzo}) implies that $X^A_+ = 0$. In turn, this makes the
left side of the second equation vanish, so the right side implies that $X^A=0$. This implies that the left side of the first equation vanishes, so the right side implies $X^A_-=0$. Thus, the only classical SUSY ground state is the trivial configuration where all $X$'s vanish.

In this derivation, we have simply analyzed the classical vacuum equations of the mass-deformed BF membrane model. One can, however, change the theory as in  \cite{Bandres:2008kj}\cite{Gomis:2008be}, and then the action we have analyzed
 is only part of a more complete BRST-invariant action, which  includes Fadeev-Popov ghosts. By using the BRST transformations of the theory, this gauge fixed action can be in  turn is related to an  even simpler classical action  \cite{Gomis:2008be}. In appendix A, we show that a similar interpretation is possible for the mass-deformed theory also. However,  in analyzing the resulting classical theory, we come to the same conclusion that the theory has no nontrivial classical vacua.\footnote{It may be that there is some alternate prescription for treating this theory where we end up with the correct physics. We note in passing that the second equation above (\ref{lorenzo}) (together with the corresponding primed equation) has, for a fixed constant $X^I_+$ a set of solutions which are in one-to-one correspondence with the solutions that we find above for the mass-deformed ABJM theory. It is not clear how to interpret this however, since the equations of motion for $X^I_+$ forbid it to have any nonzero constant value.} We should mention that there is some possibility that the single classical vacuum state that we find splits quantum mechanically into a number of distinct vacua. This is the case, for example, with vacua in the ${\cal N}=1^*$ theory. However, we have no argument that such a quantum-mechanical splitting should happen in this case. This  point  needs to be understood better and is important to clarify for the viability of the interpretation of this theory as an M2-brane theory.

\section{Discussion}

We have found that the classical vacua of the mass deformed BF membrane model does
not have the required number of vacua  to support its M2-brane interpretation.
We have performed a purely classical analysis  and it may be that a full quantum
treatment is needed. Understanding how to resolve this puzzle is important in
determining whether the maximally supersymmetric model [6][7][8] captures the
physics of M2-branes. Right now, the evidence is stacked against it.

For the mass-deformed ABJM theory, the situation seems substantially better, though still puzzling. We do find a discrete set of vacua, as expected, and we can argue that these vacua display features related to fuzzy spheres, as expected from their interpretation as giant graviton states in the Type IIB plane wave for the torus compactified theory or M2-branes expanded into M5-branes with topology $\mathbb{R}^{2,1} \times S^3_{fuzzy}$ for theory on $R^{1,2}$. However these solutions are more numerous than expected for the mass-deformed M2-brane theory. We have discussed various possible resolutions of this puzzle in section 3.1, perhaps most likely being that some of the classical vacua do not lead to genuine quantum vacua.

There are various other checks one could imagine performing directly in the $k=1$ theory. A strong check should come from the fact that there is an alternate field theory description of the DLCQ of Type IIB string theory in the plane-wave in terms of a BMN limit of a large quiver gauge theory \cite{Mukhi:2002ck}. In this alternate description, there are perturbative calculations that can be performed despite the large 't Hooft coupling since the results of such calculations obey BMN-type scaling at least at low loop order. These results should somehow be reproduced if we calculate the corresponding quantities in the ABJM theory on $T^2$, so optimistically one should be able to directly match perturbative calculations in the two theories, providing a detailed check of the proposal. It is not immediately clear how such a perturbative expansion would arise in the ABJM model, though. Related to this, there should be a BMN limit of the uncompactified theory that gives a description of M-theory on the maximally supersymmetric plane-wave. This also has an alternate description, in terms of the Plane-Wave Matrix Model, and again, one could try to match perturbative calculations between the two theories.

We have found some encouraging signs that the ABJM theory and its massive deformations contain ``fuzzy funnel" and fuzzy $S^3$'s corresponding to configurations with multiple M5-branes. This provides a new way to study the physics on multiple M5-branes by studying the physics on one of these vacua. In particular, it will be interesting to study the physics of fluctuations about the non-trivial solutions.

It is interesting to consider more generally the space of deformations of the ABJM theory, as according to the AdS/CFT or Matrix theory interpretation of the theory, these correspond to turning  on certain non-normalizable deformations near the boundary of $AdS_4\times S^7$ or to turning on linearized Type IIB supergravity backgrounds around flat spacetime
(we described two such deformations in the appendix).
In particular, one should be able to find deformations of the M2-brane theory corresponding to the whole zoo of plane-wave deformations of Type IIB string theory in flat space, preserving various amounts of supersymmetry.

%%%%%%%%%%%%%%%%%%%%%%%%%%%%%%%%%%%%%%%%%%%%%%%%%
\section*{Acknowledgments}
%%%%%%%%%%%%%%%%%%%%%%%%%%%%%%%%%%%%%%%%%%%%%%%%%%
We would like to thank Shiraz Minwalla, Filippo Passerini, Jorge Russo and Wati Taylor  for stimulating discusions, and Ofer Aharony for valuable comments on a preliminary draft. We would like to thank the Banff International Research Station and the Tata Institute for Fundamental Research, where parts of this work were completed.
Research at Perimeter Institute is supported by the Government
of Canada through Industry Canada and by the Province of Ontario through
the Ministry of Research and Innovation. J.G.  also acknowledges further  support by an NSERC Discovery Grant. The work of MVR is supported in part by the Natural Sciences and Engineering Research Council of Canada, by and Alfred P. Sloan Foundation Fellowship, and by the Canada Research Chairs Programme.D. R-G. acknowledges financial support from the European Commission through Marie Curie OIF grant contract no. MOIF-CT-2006-38381. The work of H.V. is supported in part by the National Science Foundation
under grant PHY-0243680.

%% A P P E N D I X  %%%%%%%%%%%%%%%

\setcounter{section}{0} \setcounter{subsection}{0}

\appendix{BRST-invariant mass-deformed BF-theory}

In this appendix, we consider a BRST-invariant version of the mass-deformed BF-theory theory analogous to the undeformed theory discussed in \cite{Gomis:2008uv}.

As in that paper, we can obtain a theory with no negative norm states by adding to the action of section 2 the ghost action
\begin{eqnarray}\label{LAGa}
\mathcal{L}_{\rm ghost} &=&- \partial^{\mu}c^I_- \partial_{\mu}c^I_+ + i  \bar{\chi}_+\Gamma^{\mu}\partial_{\mu}\chi_- - \mu^2 c^I_+ c^I_- + i \mu \bar{\chi}_+ \Gamma^{3456} \chi_-
\end{eqnarray}
This has the same form as the action for the $+/-$ sector, so it is also supersymmetric.

The combined action is invariant under the BRST transformations
\begin{eqnarray}
\label{brst}
\delta_{\rm brst} X_-^I =  \varepsilon \sp \ccc^I_-\, ,  & \qquad & \delta_{\rm brst} \Psi_- = \varepsilon\sp \chi_- \nonumber \\[2mm]
\delta_{\rm brst} \ccc_+^I =  \varepsilon\sp {X}_+^I\, ,  & \qquad & \delta_{\rm brst} \chi_+ = \varepsilon\sp \Psi_+.
\end{eqnarray}
Using these, it can be shown that the BRST-invariant mass-deformed theory is given by a BRST-exact term (which we can associate with gauge-fixing) plus the classical action
\be
\label{classical}
{\cal L}_0 = {\rm Tr}\Big(-\frac{1}{2}(D_{\mu}X^I )^2 - {1 \over 2} \mu^2 \tr(X^I X^I) + \frac{i}{2} \bar{\Psi}\Gamma^{\mu}D_{\mu}\Psi + {i \over 2} \bar{\Psi} \Gamma^{3546} \Psi + {1 \over 2} \epsilon^{\mu\nu\lambda}  B_{\mu} F_{\nu \lambda} \Big)\,.
\ee
There are clearly no non-trivial vacua with this action, so we reach the same conclusion as before.

Note that in the undeformed theory, one can obtain a non-trivial theory by giving $X^I_+$ a vev. The resulting theory is BRST-equivalent to a formulation of the D2-brane theory in which a formal $SO(8)$ superconformal invariance is present if we allow the $X^I_+$ vev to transform. In the mass-deformed theory, the equation of motion for $X^I_+$ becomes
\[
\partial^2 X^I_+ = \mu^2 X^I_+\,,
\]
so in this case, it is not possible to give $X^I_+$ a constant non-zero vev. It may be interesting to study the theory around a background with spacetime dependent $X^I_+$, but it is not clear what the physical interpretation would be (see \cite{Honma:2008un} for an investigation of this question).

\section{Supersymmetry transformations for the ABJM theory}

The action (\ref{ABJM}) for the ABJM theory is invariant under the following ${\cal N}=6$ SUSY variations \cite{Gaiotto:2008cg, Terashima:2008sy}:
\begin{eqnarray}
\delta C^I &=& i \omega^{IJ} \psi_J, \cr
\delta C_I^\dagger &=& i \psi^{\dagger \, J} \omega_{IJ}, \\
\delta \psi_I &=& - \gamma_\mu \omega_{IJ} D_\mu C^J
+\frac{2 \pi}{k} \left(
-\omega_{IJ} (C^K C_K^{\dagger} C^J-C^J C_c^{\dagger} C^K)
+2 \omega_{CL} C^K C_I^\dagger C^L
\right), \cr
\delta \psi^{I \dagger} &=&
D_\mu C_J^\dagger \gamma_\mu \omega^{IJ}
+\frac{2 \pi}{k} \left(
-(C_J^\dagger C^K C_K^\dagger -C_c^\dagger C^K C_J^\dagger) \omega^{IJ}
+2 C_L^\dagger C^I C_K^\dagger \omega^{CL}
\right), \\
\delta A_\mu &=& -\frac{2 \pi}{k} (
C^I \psi^{J \dagger} \gamma_\mu \omega_{IJ}
+ \omega^{IJ} \gamma_\mu \psi_I C_J^\dagger), \cr
\delta \hat{A}_\mu &=& \frac{2 \pi}{k} (
\psi^{I \dagger} C^J \gamma_\mu \omega_{IJ}
+ \omega^{IJ} \gamma_\mu C_I^\dagger \psi_J)\,.
\label{susy}
\end{eqnarray}

\section{Superfield formulations of ABJM}

The ABJM model \cite{Aharony:2008ug} can be conveniently described by using ${\cal N}=2$ superfields \cite{Benna:2008zy} or alternatively using ${\cal N}=1$ superfields  \cite{Hosomichi:2008jd} \cite{Hosomichi:2008jb}, and generalizes the
superconformal Chern-Simon construction of \cite{Gaiotto:2007qi} (see \cite{Mauri:2008ai} for an ${\cal N}=1$ superfield formulation of the Bagger-Lambert theory).
Neither of the superspace formulations makes manifest the ${\cal N}=6$ supersymmetry and corresponding $SU(4)$ R-symmetry of the theory which only becomes manifest using components fields. However, it will be useful to review them here, since the maximally supersymmetric mass-deformation that we find can be understood naturally in terms of the ${\cal N}=1$ superfield formalism (and indeed was previously written down using it), while the ${\cal N}=2$ formalism leads to other interesting mass deformations that we describe below.

\subsection{The ${\cal N}=1$ formalism}

In \cite{Hosomichi:2008jd}, the authors start with the ${\cal N}=4$  Chern-Simons theories constructed by Gaiotto and Witten \cite{Gaiotto:2008sd} and show that it is possible to add extra matter multiplets (twisted hypermultiplets) while preserving ${\cal N}=4$ supersymmetry. They further show that for the $SU(2)\times SU(2)$ theory with two bifundamental twisted hypermultiplets, one recovers the superconformal Bagger-Lambert theory. Thus, their $U(N)\times U(N)$ theory provides a natural generalization of the Bagger-Lambert theory, and one might speculate that this theory also has additional supersymmetry. In fact, we will see below that this theory is exactly the same as the theories constructed by ABJM\footnote{This was shown independently in the recent paper \cite{Hosomichi:2008jb}.}, and therefore have at least ${\cal N}=6$ supersymmetry. A nice feature of the HLLLP formalism is that there is a natural mass deformation preserving $SU(2) \times SU(2)$ supersymmetry. This was shown to reduce to the maximally supersymmetric mass-deformation of Bagger-Lambert in the $SU(2) \times SU(2)$ case, and we will see that it coincides exactly with the mass-deformation that we obtain using properties of the gravity duals to the maximally supersymmetric mass deformation.

We will describe the HLLLP theory in the special case of a $U(N) \times U(N)$ gauge group. The construction proceeds by embedding this gauge group into $Sp(2 N^2)$. We use indices $A,B$ to represent fundamental indices for this larger group, and denote the generators and invariant tensor of $Sp(2 N^2)$ by $(t^M)^A_B$ and $\omega_{AB}$ respectively. The generators corresponding to $U(N) \times U(N)$ will be denoted by $(t^m)^A_B$, and we define $k^{mn}$ as an invariant tensor for the $U(N) \times U(N)$ gauge group. The theory includes gauge fields and gauginos $(A_m)_\mu$ and $\chi_m$, together with matter multiplets $(q^A_\alpha, \psi^A_{\dot{\alpha}})$ and $(\tilde{q}^A_\alpha, \tilde{\psi}^A_{\dot{\alpha}})$ satisfying reality conditions
\[
\bar{q}_A^\alpha = \epsilon^{\alpha \beta} \omega_{AB} q^B_\beta \; , {\rm etc} \dots
\]
Below, it will be helpful to choose a more explicit representation, for which
\[
\omega = \left( \ba{cc} 0 & 1_{N^2 \times N^2} \cr -1_{N^2 \times N^2} & 0 \ea \right)
\]
and for which the generators for the left and right $U(N)$s take the form
\[
(t^{\hat{m}})^A_B = \left( \ba{cc} t^m \otimes 1 & 0 \cr 0 & t^m {}^* \otimes 1 \ea \right) \qquad (t^{m'})^A_B\left( \ba{cc} - 1 \otimes t^m  & 0 \cr 0 & -1 \otimes t^m {}^* \ea \right)
\]
where $t^m$ are the standard $U(N)$ generators. We can then write
\beas
q^A_\alpha &=& \left(\ba{c} Q_\alpha \cr -\epsilon_{\alpha \beta} Q^\beta {}^* \ea \right) \cr
\tilde{q}^A_\alpha &=& \left( \ba{c} R_\alpha \cr -\epsilon_{\alpha \beta} R^\beta {}^* \ea \right)
\eeas
where we can write $Q_\alpha$ is an $N \times N$ matrix. Then, under the gauge symmetry
\[
\left(\ba{c} Q_\alpha \cr -\epsilon_{\alpha \beta} Q^\beta {}^* \ea \right) \to \left(\ba{c} U Q_\alpha V^\dagger \cr -\epsilon_{\alpha \beta} U^* Q^\beta {}^* V^T \ea \right)\,.
\]
Finally, we can choose the invariant tensor $k^{mn}$ to have nonzero components $k^{\hat{m} \hat{n}} = \delta^{\hat{m} \hat{n}}$ and $k^{m' n'} = -\delta^{m' n'}$, which will lead to opposite signs for the two Chern-Simons terms.

The Lagrangian for the HLLLP theories are constructed in an  ${\cal N}=1$ superfield formalism with a superfield ${\cal Q}^A_\alpha$ containing the matter fields $q^A_\alpha$ and $\psi^A_{\dot{\alpha}}$ and an auxiliary field $F^A_\alpha$ and a superfield $\tilde{\cal R}^A_\beta$ for the other matter fields.

The bosonic potential arises from a superpotential
\beas
W &=& {\pi \over 6} \epsilon^{\alpha \beta} \epsilon^{\gamma \delta} k_{mn} \mu^m_{\alpha \gamma} \mu^n_{\beta \delta}  + {\pi \over 6} \epsilon^{\alpha \beta} \epsilon^{\gamma \delta} k_{mn} \tilde{\mu}^m_{\alpha \gamma} \tilde{\mu}^n_{\beta \delta} - \pi \epsilon^{\alpha \beta} \epsilon^{\gamma \delta} k_{mn} \mu^m_{\alpha \gamma} \tilde{\mu}^n_{\beta \delta} \cr
\eeas
where
\[
\mu^n_{\alpha \beta} = t^m_{AB} {\cal Q}^A_\alpha {\cal Q}^B_\beta \qquad \qquad \tilde{\mu}^n_{\alpha \beta} = t^m_{AB} {\cal R}^A_\alpha {\cal R}^B_\beta\; .
\]
Now, using the $N \times N$ matrix superfields ${\cal Q}_\alpha$ and ${\cal R}_\alpha$ introduced above, we can rewrite the superpotential in a more explicit form. We find
\bea
W &=& {\pi k \over 4} \tr({\cal Q}^\alpha {\cal Q}_\beta^\dagger {\cal Q}^\beta {\cal Q}_\alpha^\dagger - {\cal Q}^\alpha {\cal Q}_\alpha^\dagger {\cal Q}^\beta {\cal Q}_\beta^\dagger) \cr
&&+{\pi k \over 4} \tr({\cal R}^\alpha {\cal R}_\beta^\dagger {\cal R}^\beta {\cal R}_\alpha^\dagger - {\cal R}^\alpha {\cal R}_\alpha^\dagger {\cal R}^\beta {\cal R}_\beta^\dagger) \cr
&&+{\pi k \over 2} \tr(2 {\cal Q}^\alpha {\cal R}_\alpha^\dagger {\cal R}^\beta {\cal Q}_\beta^\dagger - {\cal Q}^\alpha {\cal R}_\beta^\dagger {\cal R}^\beta {\cal Q}_\alpha^\dagger - 2 {\cal Q}^\alpha {\cal Q}_\beta^\dagger {\cal R}^\beta {\cal R}_\alpha^\dagger + {\cal Q}^\alpha {\cal Q}_\alpha^\dagger {\cal R}^\beta {\cal R}_\beta^\dagger )\,.  \cr
&& 
\label{superone}
\eea
The bosonic potential is proportional to
\[
V = |M^\alpha|^2 + |N^\beta|^2
\]
where
\beas
M^\alpha &=& {\partial W \over \partial Q^\dagger_\alpha} = {\pi k \over 2}(2Q^{[\alpha} Q^\dagger_\beta Q^{\beta]} + R^{\beta} R^\dagger_\beta Q^\alpha- Q^\alpha R^\dagger_\beta R^{\beta} + 2 Q^\beta R^\dagger_\beta R^\alpha - 2R^\alpha R^\dagger_\beta  Q^\beta ) \cr
M^\alpha &=& {\partial W \over \partial R^\dagger_\alpha} =  {\pi k \over 2}(2R^{[\alpha} R^\dagger_\beta R^{\beta]} + Q^{\beta} Q^\dagger_\beta R^\alpha- R^\alpha Q^\dagger_\beta Q^{\beta} + 2 R^\beta Q^\dagger_\beta Q^\alpha - 2Q^\alpha Q^\dagger_\beta  R^\beta )\,. \cr
\eeas
While the potential in this form has only a manifest $SU(2)$ symmetry, one finds remarkably that the full potential is invariant under an $SU(2) \times SU(2)$ where the two factors act separately on $Q$ and $R$. Explicitly, we find
\beas
V &=& \tr\big(|\pi k Q^{[\alpha} Q^\dagger_\beta Q^{\beta]}|^2 + |\pi k R^{[\alpha} R^\dagger_\beta R^{\beta]}|^2 \cr
&& + \big({\pi k \over 2} \big)^2 \big(- 4 R^\alpha Q^\dagger_\gamma Q^\beta R^\dagger_\alpha Q^\gamma Q^\dagger_\beta + 2 R^\alpha Q^\dagger_\beta Q^\gamma Q^\dagger_\gamma Q^\beta R^\dagger_\alpha + 2 R^\alpha R^\dagger_\alpha Q^\beta Q^\dagger_\gamma Q^\gamma  Q^\dagger_\beta \cr
&& + 2 R^\alpha Q^\dagger_\beta Q^\beta R^\dagger_\alpha Q^\gamma Q^\dagger_\gamma - R^\alpha Q^\dagger_\beta Q^\beta Q^\dagger_\gamma Q^\gamma R^\dagger_\alpha - R^\alpha R^\dagger_\alpha Q^\beta Q^\dagger_\beta Q^\gamma Q^\dagger_\gamma \cr
&&- 4 Q^\alpha R^\dagger_\gamma R^\beta Q^\dagger_\alpha R^\gamma R^\dagger_\beta + 2 Q^\alpha R^\dagger_\beta R^\gamma R^\dagger_\gamma R^\beta Q^\dagger_\alpha + 2 Q^\alpha Q^\dagger_\alpha R^\beta R^\dagger_\gamma R^\gamma  R^\dagger_\beta \cr
&& + 2 Q^\alpha R^\dagger_\beta R^\beta Q^\dagger_\alpha R^\gamma R^\dagger_\gamma - Q^\alpha R^\dagger_\beta R^\beta R^\dagger_\gamma R^\gamma Q^\dagger_\alpha - Q^\alpha Q^\dagger_\alpha R^\beta R^\dagger_\beta R^\gamma R^\dagger_\gamma \big)\big)\,.
\eeas
Note that all terms involving indices contracted between $Q$ and $R$ have canceled. This leads to the symmetry enhancement. In fact, the potential is invariant under an $SU(4)$, since we can define $C^I = (Q^1, Q^2, R^1, R^2)$, and show that
\[
V = \big({\pi k \over 2} \big)^2 \tr(-{1 \over 3} C^I C^\dagger_I C^J C^\dagger_J C^K C^\dagger_K -{1 \over 3} C^I C^\dagger_J C^J C^\dagger_K C^K C^\dagger_I - {4 \over 3}  C^I C^\dagger_K C^J C^\dagger_I C^K C^\dagger_J + 2 C^I C^\dagger_I C^J C^\dagger_K C^K C^\dagger_J)\,.
\]
This is exactly the the bosonic potential from ABJM, so the ${\cal N}=4$ supersymmetry shared by the two theories will ensure that they are identical.

\subsubsection{Mass deformation of HLLLP}

The HLLP ${\cal N}=1$ superfield  formulation has the disadvantage that only a reduced set of the symmetries of the theory are manifest, in particular ${\cal N}=1$ supersymmetry and its corresponding $U(1)$ R-symmetry. On the other hand, since we are ultimately interested in deforming the ABJM theory, the ${\cal N}=1$ superfield formulation has the advantage that the deformations that we can add to the ABJM theory preserving at least ${\cal N}=1$ supersymmetry are completely captured by a superpotential deformation $W_{def}$. In particular, we will see that a  specific superpotential deformation proposed in \cite{Hosomichi:2008jd} corresponds to the mass-deformation that we have derived in component fields.

The mass-deformation considered in  HLLLP  \cite{Hosomichi:2008jd} involves adding to the ABJM theory   the  following superpotential deformation
\bea
W_{def} &=&  {m \over 2} \epsilon^{\alpha \beta} \omega_{AB} q^A_\alpha q^B_\beta -  {m \over 2} \epsilon^{\alpha \beta} \omega_{AB} \tilde{q}^A_\alpha \tilde{q}^B_\beta \cr
&=&+ \mu \tr({\cal Q}_\alpha {\cal Q}^\dagger_\alpha - {\cal R}_\alpha {\cal R}^\dagger_\alpha),
\label{defoo}
\eea
where $\mu$ is proportional to $m$. Now, the bosonic potential of the deformed theory is proportional to
\bea
V = |M^\alpha|^2 + |N^\beta|^2,
\label{poten}
\eea
where
\beas
M^\alpha &=& {\partial W \over \partial Q^\dagger_\alpha} = \mu Q^\alpha + {\pi k \over 2}(2Q^{[\alpha} Q^\dagger_\beta Q^{\beta]} + R^{\beta} R^\dagger_\beta Q^\alpha- Q^\alpha R^\dagger_\beta R^{\beta} + 2 Q^\beta R^\dagger_\beta R^\alpha - 2R^\alpha R^\dagger_\beta  Q^\beta ) \cr
N^\alpha &=& {\partial W \over \partial R^\dagger_\alpha} = - \mu R^\alpha + {\pi k \over 2}(2R^{[\alpha} R^\dagger_\beta R^{\beta]} + Q^{\beta} Q^\dagger_\beta R^\alpha- R^\alpha Q^\dagger_\beta Q^{\beta} + 2 R^\beta Q^\dagger_\beta Q^\alpha - 2Q^\alpha Q^\dagger_\beta  R^\beta ).
\eeas
Here $W$ combines the original ABJM superpotential (\ref{superone}) with the deformed contribution (\ref{defoo})

We see that this is exactly the deformation we arrived at in our component analysis.

\subsection{The ${\cal N}=2$ formalism.}

The ABJM theory may also be written in terms of an ${\cal N}=2$ superfield formalism \cite{Benna:2008zy}, which leads to a couple of additional interesting mass deformations that we discuss below. In terms of ${\cal N}=2$ superfields, the  ABJM theory consists of two vector superfields $\VV$, $\hat{\VV}$  and chiral superfields $\ZZ_a,\bar{\ZZ}_a,\WW_{\dot{a}},\bar{\WW}_{\dot{a}}$, which tranform in the following representations of the $U(N)\times U(N)$ gauge group.

The ABJM theory is described by the following action
\bea
S=S_{CS}+S_{mat}+S_{pot},
\eea
where
\bea
 S_{CS}&=& -i k \int d^3x\,d^4\theta \int_0^1 dt\: \tr \left[ \VV \bar{D}^\alpha \left( e^{t \VV} D_\alpha e^{-t \VV} \right) - \hat{\VV} \bar{D}^\alpha \left( e^{t \hat{\VV}} D_\alpha e^{-t \hat{\VV}}  \right)\right] \; , \label{CS-action} \\
 S_{mat}&=&- \int d^3x\,d^4\theta\: \tr \left(\bar{\ZZ}_a e^{-\VV} \ZZ_a e^{\hat{\VV}}+
 \bar{\WW}_{\dot{a}} e^{-\hat{\VV}} \WW_{\dot{a}} e^{{\VV}}\right)\; , \label{mat-action-2-2} \\
  S_{pot}&=& {1\over k}  \int d^3x\,d^2\theta\: W(\ZZ,\WW)
   + {1\over k} \int d^3x\,d^2\bar{\theta}\: \bar{W}(\bar{\ZZ},\bar{\WW})
 \label{superpot-2-2}
\eea
with
\be  W    = \frac{1}{4} \epsilon^{ab}\epsilon^{\dot{a}\dot{b}} \tr \ZZ_a \WW_{\dot{a}} \ZZ_b \WW_{\dot{b}}
  \qquad  \bar{W} = \frac{1}{4}\epsilon^{ab}\epsilon^{\dot{a}\dot{b}} \tr \bar{\ZZ}_a \bar{\WW}_{\dot{a}}  \bar{\ZZ}_b \bar{\WW}_{\dot{b}}
  \; .
\ee
The scalar potential of the theory has two contributions
\bea
V_{scalar}=V_{super}+V_{CS}.
\eea
$V_{super}$ is the usual contribution to the potential from the superpotential of the theory (\ref{superpot-2-2})
\bea
V_{super}=\tr\left(F^aF^{a\dagger}+G^{\dot{a}}G^{\dot{a}\dagger}\right)\geq 0,
\eea
where
\bea
F^{a\dagger}={1\over 2k}\epsilon^{ab}\epsilon^{\dot{a}\dot{b}} \WW_{\dot{a}} \ZZ_b \WW_{\dot{b}}\qquad
G^{a\dagger}=-{1\over 2k}\epsilon^{ab}\epsilon^{\dot{a}\dot{b}} \ZZ_a\WW_{\dot{b}} \ZZ_b
\eea
The scalar potential piece $V_{CS}$ arises by integrating out the $D$ and $\hat{D}$ term in the vector multiplets $\VV$ and
$\hat{\VV}$ , which in turn constraint the scalar fields in  $\VV$, $\hat{\VV}$ to take the following values
\bea
\sigma={1\over 4k}\left(A_aA_a^\dagger-B_{\dot{a}}^\dagger B_{\dot{a}}\right)\qquad \hat{\sigma}={1\over 4k}\left(A_a^\dagger A_a- B_{\dot{a}} B_{\dot{a}}^\dagger \right).
\label{constraint}
\eea
As shown in \cite{Benna:2008zy}, the following terms generate the scalar potential arising from the vector multiplets $\VV$, $\hat{\VV}$
is given by
\bea
V_{CS}=\tr\left(A_a^\dagger \sigma^2 A_a
   +A_a^\dagger A_a \hat{\sigma}^2
   - 2A_a^\dagger \sigma A_a \hat{\sigma}\right)+\tr\left(  B_{\dot{a}}^\dagger \hat{\sigma}^2 B_{\dot{a}}+
   B_{\dot{a}}^\dagger B_{\dot{a}} \sigma^2
   - 2B_{\dot{a}}^\dagger \hat{\sigma} B_{\dot{a}} \sigma\right),
 \eea
where $\sigma$ and $\hat{\sigma}$ are given by (\ref{constraint}). We can now further rewrite the potential as
\bea
V_{CS}=\tr\left|\sigma A_a-A_a \hat{\sigma}\right|^2+\tr\left|\hat{\sigma} B_{\dot{a}}-B_{\dot{a}} \sigma\right|^2\geq 0,
\eea
which makes manifest the positivity of the scalar potential.

In the present description, the manifest global symmetry of the theory is $SU(2) \times SU(2) \times U(1)$. However, as shown in \cite{Benna:2008zy}, the theory actually has a manifest $SU(4)$ R-symmetry. For example, if we define $C^I = (A^1, A^2, B^{\dagger 1}, B^{\dagger 2})$, we may rewrite the bosonic potential as
\[
V = \big({2\pi  \over k} \big)^2 \tr(-{1 \over 3} C^I C^\dagger_I C^J C^\dagger_J C^K C^\dagger_K -{1 \over 3} C^I C^\dagger_J C^J C^\dagger_K C^K C^\dagger_I - {4 \over 3}  C^I C^\dagger_K C^J C^\dagger_I C^K C^\dagger_J + 2 C^I C^\dagger_I C^J C^\dagger_K C^K C^\dagger_J)\,.
\]
which is exactly the expression we had in the component action.

\section{Other supersymmetric mass deformations of ABJM}

In this section, we consider a couple of other interesting mass deformations of ABJM, which appear naturally in the ${\cal N}=2$ superfield formalism . In the ${\cal N}=2$ superfield formulation, the supersymmetric deformations of the action are encoded by a superpotential deformation  $W_{def}$ {\it and} a D-term deformation (a different deformation was considered in \cite{Benna:2008zy}. We will consider deformations of both types now.

\subsection{FI deformation}

We first consider the D-term deformation, given by following action
\bea
S=S_{CS}+S_{mat}+S_{pot}+S_{FI},
\label{totall}
\eea
with
\bea
S_{FI}= k \int d^3x\,d^4\theta\,{\mu}\tr\left(\VV+\hat{\VV}\right)
\label{FI}
\eea
and where $S_{CS}$,$S_{mat}$ and $S_{pot}$ are given respectively in (\ref{CS-action})(\ref{mat-action-2-2})(\ref{superpot-2-2}). This corresponds to adding the following $D$-terms into the component action
\bea
S_{FI}= k \int d^3x\,{\mu}\tr (D+\hat{D}).
\eea
We now analyze the effect of this deformation and explore the effect it has on the theory.

An unusual property of the superfield formulation of the ABJM action is that the two scalar fields of the vector multiplet $\VV/\hat{\VV}$ in the WZ gauge -- given by $\sigma/\hat{\sigma}$ and $D/\hat{D}$ -- are auxilliary fields. The origin of this is the peculiar form of the action for the gauge fields in (\ref{CS-action}). In writing down the theory in components we have to integrate out sequentially  $D/\hat{D}$  and $\sigma/\hat{\sigma}$.

When the action in (\ref{totall}) is expanded in components one finds that $D$ and $\hat{D}$ appears at most linearly, unlike in conventional gauge theories, where it appears quadratically. In the case of the ABJM theory, varying the action for $D$ and $\hat{D}$ identifies $\sigma$ and $\hat{\sigma}$ in terms of the matter fields (\ref{constraint}). The effect of the deformation (\ref{FI}) is therefore to shift the value of $\sigma$ and $\hat{\sigma}$ by the mass parameter $\mu$
\bea
\sigma\rightarrow \sigma +{\mu\over 4}\cr
\cr
\hat{\sigma}\rightarrow \hat{\sigma} -{\mu\over 4}\cr
\eea

Due to this shift induced on the scalar fields in the vector multiplet, the scalar potential $V_{CS}$ coming from the vector multiplets  is now given by
\bea
V_{CS}=\tr\left|{\mu\over 2}  A_a+\sigma A_a-A_a \hat{\sigma}\right|^2+\tr\left|{\mu\over 2}B_{\dot{a}} +B_{\dot{a}} \sigma-\hat{\sigma} B_{\dot{a}}\right|^2\geq 0,
\eea
where for completeness we write  that
\bea
\sigma={1\over 4k}\left(A_aA_a^\dagger-B_{\dot{a}}^\dagger B_{\dot{a}}\right)\qquad \hat{\sigma}={1\over 4k}\left(A_a^\dagger A_a- B_{\dot{a}} B_{\dot{a}}^\dagger \right).
\label{constraintma}
\eea

Therefore, one effect of the mass deformation is to give mass to the matter fields, both to the scalars and the fermions. The mass terms for the matter fields are
\bea
{\cal L}_{mass}={\mu^2\over 4}\tr\left(A_aA_a^\dagger+B_{\dot{a}}^\dagger B_{\dot{a}}\right)+i{\mu\over 2}\tr\left(\zeta^\dagger_a\zeta_a-
   \omega_{\dot{a}}^\dagger \omega_{\dot{a}} \right)\,.
\eea
The other effect of our deformation is to turn on new terms in the scalar potential, which can be interpreted as arising from a background flux
\beas
{\cal L}_{flux}={\mu\over 2}\tr\left( A_a(\bar{\sigma}{\bar A}_a-{\bar A}_a \bar{\hat{\sigma}})+
\bar{A}_a(\bar{\sigma}{ A}_a-{\bar A}_a {\hat{\sigma}})+B_{\dot{a}}(\bar{B}_{\dot{a}} \bar{\sigma}-\bar{\hat{\sigma}}\bar{B}_{\dot{a}}+\bar{B}_{\dot{a}}({B}_{\dot{a}} \bar{\sigma}-{\hat{\sigma}}{B}_{\dot{a}})\right)
\eeas

\subsubsection{Vacua of the FI-deformed theory}

The bosonic potential for the deformed theory may be written as a sum of squares
\beas
V &=& |{4 \pi \over k} \epsilon_{ab} \epsilon_{\dot{a} \dot{b}} B_{\dot{a}} A_b B_{\dot{b}}|^2 + |{4 \pi \over k} \epsilon_{ab} \epsilon_{\dot{a} \dot{b}} A_b B_{\dot{b}} A_a |^2 \cr
&&+|\mu A_c + \sigma_1 A_c - A_c \sigma_2|^2 + |\mu B_{\dot{c}} + \sigma_2 B_{\dot{c}} - B_{\dot{c}} \sigma_1|^2\,.
\eeas
Demanding that the potential vanishes gives
\beas
\epsilon_{ab} \epsilon_{\dot{a} \dot{b}} B_{\dot{a}} A_b B_{\dot{b}} &=& 0 \cr
\epsilon_{ab} \epsilon_{\dot{a} \dot{b}} A_b B_{\dot{b}} A_a  &=& 0 \cr
\mu A_c + \sigma_1 A_c - A_c \sigma_2 &=& 0 \cr
\mu B_{\dot{c}} + \sigma_2 B_{\dot{c}} - B_{\dot{c}} \sigma_1 &=& 0\,.
\eeas
The modification breaks the $SU(4)$ symmetry (which is not manifest in this description) to the manifest $SU(2) \times SU(2)$ symmetry. At this point, we note that for the choice $B=0$, the equations reduce to
\bea
\label{eqns2}
A_1 + A_2 A_2^\dagger A_1 - A_1 A_2^\dagger A_2 &=& 0 \cr
A_2 + A_1 A_1^\dagger A_2 - A_2 A_1^\dagger A_1 &=& 0
\eea
which are exactly the ones we analyzed in section 4. Similarly, we can set $A=0$ to get a set of equations that may be obtained from the $A$ equations by the replacement $A \to B^\dagger$. Thus, this deformation also has discrete solutions in one-to-one correspondence with partitions of $N$, where the elements of the partition of size greater than one are labeled either by $A$ or $B$.

\subsection{Another superpotential deformation of ABJM.}

We now consider a third deformation of ABJM, obtained by adding a superpotential deformation
\[
\delta W = \mu \tr(\epsilon^{a \dot{b}} A_a B_{\dot{b}})\,.
\]
This apparently breaks $SU(2) \times SU(2)$ down to a diagonal $SU(2)$. With this deformation, the bosonic potential vanishes if and only if
\beas
\mu \epsilon^{a \dot{a}} B_{\dot{a}}  + \epsilon^{ab} \epsilon^{\dot{a} \dot{b}} B_{\dot{a}} A_b B_{\dot{b}} &=& 0 \cr
\mu \epsilon^{a \dot{a}} A_a  + \epsilon^{ab} \epsilon^{\dot{a} \dot{b}} A_b B_{\dot{b}} A_a  &=& 0 \cr
\sigma_1 A_c - A_c \sigma_2 &=& 0 \cr
\sigma_2 B_{\dot{c}} - B_{\dot{c}} \sigma_1 &=& 0\,.
\eeas
Again, we will not give a complete analysis of these equations, but simply write down a set of solutions. We note that for
\[
B_\alpha = \pm \epsilon_{\alpha \beta} A^{\dagger \beta}
\]
we have $\sigma_1=\sigma_2=0$, so the third and fourth equation are automatically satisfied, while the other equations are satisfied if and only if
\[
A_\alpha \pm A_\beta A^{\dagger \beta} A_\alpha \mp A_\alpha A^{\dagger \beta} A_\beta\,.
\]
But these are the same as the equations (\ref{eqns2}) that arose in the other mass deformations, so again, we have (at least) two irreducible solutions for each $N$.
\vfill\eject


\begin{thebibliography}{999}

%\cite{Bagger:2006sk}
\bibitem{Bagger:2006sk}
  J.~Bagger and N.~Lambert,
  ``Modeling multiple M2's,''
  Phys.\ Rev.\  D {\bf 75}, 045020 (2007)
  [arXiv:hep-th/0611108].
  %%CITATION = PHRVA,D75,045020;%%


%\cite{Bagger:2007jr}
\bibitem{Bagger:2007jr}
  J.~Bagger and N.~Lambert,
  ``Gauge Symmetry and Supersymmetry of Multiple M2-Branes,''
  Phys.\ Rev.\  D {\bf 77}, 065008 (2008)
  [arXiv:0711.0955 [hep-th]].
  %%CITATION = PHRVA,D77,065008;%%

%\cite{Bagger:2007vi}
\bibitem{Bagger:2007vi}
  J.~Bagger and N.~Lambert,
  ``Comments On Multiple M2-branes,''
  JHEP {\bf 0802}, 105 (2008)
  [arXiv:0712.3738 [hep-th]].
  %%CITATION = JHEPA,0802,105;%%


%\cite{Gustavsson:2007vu}
\bibitem{Gustavsson:2007vu}
  A.~Gustavsson,
  ``Algebraic structures on parallel M2-branes,''
  arXiv:0709.1260 [hep-th].
  %%CITATION = ARXIV:0709.1260;%%

%\cite{Gustavsson:2008dy}
\bibitem{Gustavsson:2008dy}
  A.~Gustavsson,
  ``Selfdual strings and loop space Nahm equations,''
  JHEP {\bf 0804}, 083 (2008)
  [arXiv:0802.3456 [hep-th]].
  %%CITATION = JHEPA,0804,083;%%

%\cite{Gomis:2008uv}
\bibitem{Gomis:2008uv}
  J.~Gomis, G.~Milanesi and J.~G.~Russo,
  ``Bagger-Lambert Theory for General Lie Algebras,''
  arXiv:0805.1012 [hep-th].
  %%CITATION = ARXIV:0805.1012;%%

%\cite{Benvenuti:2008bt}
\bibitem{Benvenuti:2008bt}
  S.~Benvenuti, D.~Rodriguez-Gomez, E.~Tonni and H.~Verlinde,
  ``N=8 superconformal gauge theories and M2 branes,''
  arXiv:0805.1087 [hep-th].
  %%CITATION = ARXIV:0805.1087;%%

%\cite{Ho:2008ei}
\bibitem{Ho:2008ei}
  P.~M.~Ho, Y.~Imamura and Y.~Matsuo,
  ``M2 to D2 revisited,''
  arXiv:0805.1202 [hep-th].
  %%CITATION = ARXIV:0805.1202;%%

%\cite{Honma:2008jd}
\bibitem{Honma:2008jd}
  Y.~Honma, S.~Iso, Y.~Sumitomo and S.~Zhang,
  ``Scaling limit of N=6 superconformal Chern-Simons theories and Lorentzian
  Bagger-Lambert theories,''
  arXiv:0806.3498 [hep-th].
  %%CITATION = ARXIV:0806.3498;%%


\bibitem{Banerjee:2008pd}
  S.~Banerjee and A.~Sen,
  ``Interpreting the M2-brane Action,''
  arXiv:0805.3930 [hep-th].
  %%CITATION = ARXIV:0805.3930;%%

%\cite{Cecotti:2008qs}
\bibitem{Cecotti:2008qs}
  S.~Cecotti and A.~Sen,
  ``Coulomb Branch of the Lorentzian Three Algebra Theory,''
  arXiv:0806.1990 [hep-th].
  %%CITATION = ARXIV:0806.1990;%%

  %\cite{Bandres:2008kj}
\bibitem{Bandres:2008kj}
  M.~A.~Bandres, A.~E.~Lipstein and J.~H.~Schwarz,
  ``Ghost-Free Superconformal Action for Multiple M2-Branes,''
  arXiv:0806.0054 [hep-th].
  %%CITATION = ARXIV:0806.0054;%%



  %\cite{Gomis:2008be}
\bibitem{Gomis:2008be}
  J.~Gomis, D.~Rodriguez-Gomez, M.~Van Raamsdonk and H.~Verlinde,
  ``The Superconformal Gauge Theory on M2-Branes,''
  arXiv:0806.0738 [hep-th].
  %%CITATION = ARXIV:0806.0738;%%

%\cite{Ezhuthachan:2008ch}
\bibitem{Ezhuthachan:2008ch}
  B.~Ezhuthachan, S.~Mukhi and C.~Papageorgakis,
  ``D2 to D2,''
  arXiv:0806.1639 [hep-th].
  %%CITATION = ARXIV:0806.1639;%%

  %\cite{Aharony:2008ug}
\bibitem{Aharony:2008ug}
  O.~Aharony, O.~Bergman, D.~L.~Jafferis and J.~Maldacena,
  ``N=6 superconformal Chern-Simons-matter theories, M2-branes and their
  gravity duals,''
  arXiv:0806.1218 [hep-th].
  %%CITATION = ARXIV:0806.1218;%%

%\cite{Lin:2004nb}
\bibitem{Lin:2004nb}
  H.~Lin, O.~Lunin and J.~M.~Maldacena,
  ``Bubbling AdS space and 1/2 BPS geometries,''
  JHEP {\bf 0410}, 025 (2004)
  [arXiv:hep-th/0409174].
  %%CITATION = JHEPA,0410,025;%%

\bibitem{Bhattacharya:2008bj}
  J.~Bhattacharya and S.~Minwalla,
  ``Superconformal Indices for ${\cal N}=6$ Chern Simons Theories,''
  arXiv:0806.3251 [hep-th].
  %%CITATION = ARXIV:0806.3251;%%

%\cite{Gomis:2008cv}
\bibitem{Gomis:2008cv}
  J.~Gomis, A.~J.~Salim and F.~Passerini,
  ``Matrix Theory of Type IIB Plane Wave from Membranes,''
  arXiv:0804.2186 [hep-th].
  %%CITATION = ARXIV:0804.2186;%%

%\cite{Hosomichi:2008qk}
\bibitem{Hosomichi:2008qk}
  K.~Hosomichi, K.~M.~Lee and S.~Lee,
  ``Mass-Deformed Bagger-Lambert Theory and its BPS Objects,''
  arXiv:0804.2519 [hep-th].
  %%CITATION = ARXIV:0804.2519;%%
  %\cite{Mukhi:2002ck}
\bibitem{Mukhi:2002ck}
  S.~Mukhi, M.~Rangamani and E.~P.~Verlinde,
  ``Strings from quivers, membranes from moose,''
  JHEP {\bf 0205}, 023 (2002)
  [arXiv:hep-th/0204147].
  %%CITATION = JHEPA,0205,023;%%


\if 0
%\cite{Bena:2000zb}
\bibitem{Bena:2000zb}
  I.~Bena,
  ``The M-theory dual of a 3 dimensional theory with reduced supersymmetry,''
  Phys.\ Rev.\  D {\bf 62}, 126006 (2000)
  [arXiv:hep-th/0004142].
  %%CITATION = PHRVA,D62,126006;%%
\fi

%\cite{VanRaamsdonk:2008ft}
\bibitem{VanRaamsdonk:2008ft}
  M.~Van Raamsdonk,
  ``Comments on the Bagger-Lambert theory and multiple M2-branes,''
  JHEP {\bf 0805}, 105 (2008)
  [arXiv:0803.3803 [hep-th]].
  %%CITATION = JHEPA,0805,105;%%


%%\cite{Benna:2008zy}
\bibitem{Benna:2008zy}
  M.~Benna, I.~Klebanov, T.~Klose and M.~Smedback,
  ``Superconformal Chern-Simons Theories and $AdS_4/CFT_3$ Correspondence,''
  arXiv:0806.1519 [hep-th].
  %%CITATION = ARXIV:0806.1519;%%
\bibitem{Hosomichi:2008jd}
  K.~Hosomichi, K.~M.~Lee, S.~Lee, S.~Lee and J.~Park,
  ``N=4 Superconformal Chern-Simons Theories with Hyper and Twisted Hyper
  Multiplets,''
  arXiv:0805.3662 [hep-th].
  %%CITATION = ARXIV:0805.3662;%%

  %\cite{Hosomichi:2008jb}
\bibitem{Hosomichi:2008jb}
  K.~Hosomichi, K.~M.~Lee, S.~Lee, S.~Lee and J.~Park,
  ``N=5,6 Superconformal Chern-Simons Theories and M2-branes on Orbifolds,''
  arXiv:0806.4977 [hep-th].
  %%CITATION = ARXIV:0806.4977;%%

  %\cite{Basu:2004ed}
\bibitem{Basu:2004ed}
  A.~Basu and J.~A.~Harvey,
   ``The M2-M5 brane system and a generalized Nahm's equation,''
  Nucl.\ Phys.\  B {\bf 713}, 136 (2005)
  [arXiv:hep-th/0412310].
  %%CITATION = NUPHA,B713,136;%%

    %\cite{Gaiotto:2008sd}
\bibitem{Gaiotto:2008sd}
  D.~Gaiotto and E.~Witten,
  ``Janus Configurations, Chern-Simons Couplings, And The Theta-Angle in N=4
  Super Yang-Mills Theory,''
  arXiv:0804.2907 [hep-th].
  %%CITATION = ARXIV:0804.2907;%%

    %\cite{Constable:1999ac}
\bibitem{Constable:1999ac}
  N.~R.~Constable, R.~C.~Myers and O.~Tafjord,
   ``The noncommutative bion core,''
  Phys.\ Rev.\  D {\bf 61}, 106009 (2000)
  [arXiv:hep-th/9911136].
  %%CITATION = PHRVA,D61,106009;%%


%\cite{Honma:2008un}
\bibitem{Honma:2008un}
  Y.~Honma, S.~Iso, Y.~Sumitomo and S.~Zhang,
  ``Janus field theories from multiple M2 branes,''
  arXiv:0805.1895 [hep-th].
  %%CITATION = ARXIV:0805.1895;%%
  
\bibitem{Gaiotto:2008cg}
  D.~Gaiotto, S.~Giombi and X.~Yin,
  ``Spin Chains in N=6 Superconformal Chern-Simons-Matter Theory,''
  arXiv:0806.4589 [hep-th].
  %%CITATION = ARXIV:0806.4589;%%

%\cite{Terashima:2008sy}
\bibitem{Terashima:2008sy}
  S.~Terashima,
  ``On M5-branes in N=6 Membrane Action,''
  arXiv:0807.0197 [hep-th].
  %%CITATION = ARXIV:0807.0197;%%

    %\cite{Bena:2004jw}
\bibitem{Bena:2004jw}
  I.~Bena and N.~P.~Warner,
  ``A harmonic family of dielectric flow solutions with maximal
  supersymmetry,''
  JHEP {\bf 0412}, 021 (2004)
  [arXiv:hep-th/0406145].
  %%CITATION = JHEPA,0412,021;%%

%\cite{Bena:2000zb}
\bibitem{Bena:2000zb}
  I.~Bena,
  ``The M-theory dual of a 3 dimensional theory with reduced supersymmetry,''
  Phys.\ Rev.\  D {\bf 62}, 126006 (2000)
  [arXiv:hep-th/0004142].
  %%CITATION = PHRVA,D62,126006;%%

\bibitem{Polchinski:2000uf}
  J.~Polchinski and M.~J.~Strassler,
  ``The string dual of a confining four-dimensional gauge theory,''
  arXiv:hep-th/0003136.
  %%CITATION = HEP-TH/0003136;%%

%\cite{Banks:1996vh}
\bibitem{Banks:1996vh}
  T.~Banks, W.~Fischler, S.~H.~Shenker and L.~Susskind,
  ``M theory as a matrix model: A conjecture,''
  Phys.\ Rev.\  D {\bf 55}, 5112 (1997)
  [arXiv:hep-th/9610043].
  %%CITATION = PHRVA,D55,5112;%%



\bibitem{Taylor:1998tv}
  W.~Taylor and M.~Van Raamsdonk,
  ``Supergravity currents and linearized interactions for matrix theory
  configurations with fermionic backgrounds,''
  JHEP {\bf 9904}, 013 (1999)
  [arXiv:hep-th/9812239].
  %%CITATION = JHEPA,9904,013;%%

%\cite{SheikhJabbari:2004ik}
\bibitem{SheikhJabbari:2004ik}
  M.~M.~Sheikh-Jabbari,
  ``Tiny graviton matrix theory: DLCQ of IIB plane-wave string theory, a
  conjecture,''
  JHEP {\bf 0409}, 017 (2004)
  [arXiv:hep-th/0406214].
  %%CITATION = JHEPA,0409,017;%%
%\cite{Lozano:2006jr}
\bibitem{Lozano:2006jr}
  Y.~Lozano and D.~Rodriguez-Gomez,
  ``Type II pp-wave matrix models from point-like gravitons,''
  JHEP {\bf 0608}, 022 (2006)
  [arXiv:hep-th/0606057].
  %%CITATION = JHEPA,0608,022;%%

\bibitem{Bagger:2008se}
  J.~Bagger and N.~Lambert,
  ``Three-Algebras and N=6 Chern-Simons Gauge Theories,''
  arXiv:0807.0163 [hep-th].
  %%CITATION = ARXIV:0807.0163;%%
%\cite{Maldacena:1997re}
\bibitem{Maldacena:1997re}
  J.~M.~Maldacena,
  ``The large N limit of superconformal field theories and supergravity,''
  Adv.\ Theor.\ Math.\ Phys.\  {\bf 2}, 231 (1998)
  [Int.\ J.\ Theor.\ Phys.\  {\bf 38}, 1113 (1999)]
  [arXiv:hep-th/9711200].
  %%CITATION = IJTPB,38,1113;%%
  %\cite{Gubser:1998bc}
\bibitem{Gubser:1998bc}
  S.~S.~Gubser, I.~R.~Klebanov and A.~M.~Polyakov,
  ``Gauge theory correlators from non-critical string theory,''
  Phys.\ Lett.\  B {\bf 428}, 105 (1998)
  [arXiv:hep-th/9802109].
  %%CITATION = PHLTA,B428,105;%%

%\cite{Witten:1998qj}
\bibitem{Witten:1998qj}
  E.~Witten,
  ``Anti-de Sitter space and holography,''
  Adv.\ Theor.\ Math.\ Phys.\  {\bf 2}, 253 (1998)
  [arXiv:hep-th/9802150].
  %%CITATION = 00203,2,253;%%

%\cite{Sethi:1997sw}
\bibitem{Sethi:1997sw}
  S.~Sethi and L.~Susskind,
  ``Rotational invariance in the M(atrix) formulation of type IIB theory,''
  Phys.\ Lett.\  B {\bf 400}, 265 (1997)
  [arXiv:hep-th/9702101].
  %%CITATION = PHLTA,B400,265;%%

%\cite{Banks:1996my}
\bibitem{Banks:1996my}
  T.~Banks and N.~Seiberg,
  ``Strings from matrices,''
  Nucl.\ Phys.\  B {\bf 497}, 41 (1997)
  [arXiv:hep-th/9702187].
  %%CITATION = NUPHA,B497,41;%%

%\cite{Berenstein:2002jq}
\bibitem{Berenstein:2002jq}
  D.~E.~Berenstein, J.~M.~Maldacena and H.~S.~Nastase,
  ``Strings in flat space and pp waves from N = 4 super Yang Mills,''
  JHEP {\bf 0204}, 013 (2002)
  [arXiv:hep-th/0202021].
  %%CITATION = JHEPA,0204,013;%%

%\cite{Dasgupta:2002hx}
\bibitem{Dasgupta:2002hx}
  K.~Dasgupta, M.~M.~Sheikh-Jabbari and M.~Van Raamsdonk,
  ``Matrix perturbation theory for M-theory on a PP-wave,''
  JHEP {\bf 0205}, 056 (2002)
  [arXiv:hep-th/0205185].
  %%CITATION = JHEPA,0205,056;%%


%\cite{Schwarz:2004yj}
\bibitem{Schwarz:2004yj}
  J.~H.~Schwarz,
  ``Superconformal Chern-Simons theories,''
  JHEP {\bf 0411}, 078 (2004)
  [arXiv:hep-th/0411077].
  %%CITATION = JHEPA,0411,078;%%
  %\cite{Nahm:1977tg}
\bibitem{Nahm:1977tg}
  W.~Nahm,
  ``Supersymmetries and their representations,''
  Nucl.\ Phys.\  B {\bf 135}, 149 (1978).
  %%CITATION = NUPHA,B135,149;%%
  %\cite{Lin:2005nh}
\bibitem{Lin:2005nh}
  H.~Lin and J.~M.~Maldacena,
  ``Fivebranes from gauge theory,''
  Phys.\ Rev.\  D {\bf 74}, 084014 (2006)
  [arXiv:hep-th/0509235].
  %%CITATION = PHRVA,D74,084014;%%

\bibitem{Gaiotto:2007qi}
  D.~Gaiotto and X.~Yin,
  ``Notes on superconformal Chern-Simons-matter theories,''
  JHEP {\bf 0708}, 056 (2007)
  [arXiv:0704.3740 [hep-th]].
  %%CITATION = JHEPA,0708,056;%%

%\cite{Mauri:2008ai}
\bibitem{Mauri:2008ai}
  A.~Mauri and A.~C.~Petkou,
  ``An N=1 Superfield Action for M2 branes,''
  arXiv:0806.2270 [hep-th].
  %%CITATION = ARXIV:0806.2270;%%


\end{thebibliography}
\end{document}